\documentclass[10pt,twocolumn,twoside]{IEEEtran}
\IEEEoverridecommandlockouts
\usepackage{cite}
\usepackage[mathscr]{eucal}
\usepackage{amsmath,amssymb,amsfonts, mathtools}
\usepackage{graphicx}
\usepackage{textcomp}
\usepackage{amsthm}
\usepackage{array}
\usepackage{hyperref}
\usepackage{dsfont}
\usepackage{bm}
\usepackage{caption}
\usepackage{soul}
\usepackage{color}
\usepackage{subcaption}
\usepackage{enumitem}
\usepackage{booktabs}
\usepackage{multirow}
\usepackage{algorithm}
\usepackage{algpseudocode}
\algrenewcommand\algorithmicrequire{\textbf{Input:}}
\algrenewcommand\algorithmicensure{\textbf{Output:}}
\usepackage{float}
\usepackage{subfiles}
\usepackage{verbatim}
\usepackage{bbm}

\newtheorem{theorem}{Theorem}
\newtheorem{remark}{Remark}

\usepackage{stfloats}
\newtheorem{proposition}{Proposition}
\newtheorem{lemma}{Lemma}
\newcommand{\cT}{\mathcal T}
\newcommand{\pibf}{\bm \pi}
\algrenewcommand\algorithmicindent{0.5em}
\usepackage{xcolor}
\def\BibTeX{{\rm B\kern-.05em{\sc i\kern-.025em b}\kern-.08em
    T\kern-.1667em\lower.7ex\hbox{E}\kern-.125emX}}

\def\ie{{\it i.e.,\ \/}}
\def\eg{{\it e.g., \/}}
\def\etal{{\it et~al. \/}}

\begin{document}

\title{
Joint Scheduling of Deferrable and Nondeferrable Demand with Colocated Stochastic Supply}
\author{Minjae~Jeon,~\IEEEmembership{Student Member,~IEEE,}
        Lang~Tong,~\IEEEmembership{Fellow,~IEEE,}
        Qing~Zhao,~\IEEEmembership{Fellow,~IEEE}
    \thanks{Minjae Jeon, Lang Tong, and Qing Zhao(\{\textcolor{blue}{\texttt{mj444, lt35, qz16}}\}\textcolor{blue}{\texttt{@cornell.edu}}) are with the School of Electrical and Computer Engineering, Cornell University, USA. This work was supported in part by the National Science Foundation under Grant 2412776  and 2419622. A small part of this work was presented at the 62nd IEEE
Conf. Decision and Control (CDC ‘23).}}

\maketitle
\begin{abstract}
We investigate the problem of serving deferrable and nondeferrable electric demands with colocated stochastic supply and grid-imported electricity. Deferrable demands arrive randomly and can be delayed within their service deadlines. Nondeferrable demands are always present and must be served immediately, but the quantity served depends on the cost of electricity. Colocated supply is stochastic with zero marginal cost. It can be used to meet demand or exported to the grid to maximize profit. The stochasticity of demands and local supply makes optimal scheduling a Markov decision process with continuous (uncountable) state and action spaces. Under deterministic, time-varying, and piecewise-linear retail pricing of electricity, we show that the optimal demand scheduling follows the {\em Principle of Procrastination}, which reduces the infinite-dimensional policy space to a finite-dimensional Euclidean space defined by three procrastination parameters for each deferrable demand. For settings in which the underlying probability distributions are unknown, we propose a {\em Procrastination Threshold Reinforcement Learning} algorithm. Numerical experiments based on real-world test data confirm that the proposed threshold learning algorithm closely approximates the optimal policy and outperforms standard benchmarks.

\end{abstract}

\section{Introduction}
The classic problem of scheduling demands (jobs) with deadlines dates back to the late 1950s and 60s. McNaughton \cite{mcnaughton1959scheduling} considered the scheduling of jobs with ``soft deadlines,’’ where a penalty is imposed if the job is not completed by the deadline. The hard deadline scheduling problem for real-time multiprocessor control was studied by Manacher \cite{manacher1967production}. Since the seminal work of Liu and Layland \cite{liu1973scheduling}, an extensive theory of deadline scheduling has been developed. See the historical surveys by Sha \etal \cite{sha2004real} and by Davis and Burns \cite{davis2011survey}. The application domain of deadline scheduling is broad and continues to expand to control system design \cite{seidel2026controller}, time-sensitive networking \cite{jaramillo2011scheduling, stuber2023survey}, electric vehicle (EV) charging \cite{chen2012deadline,gan2012optimal}, and data centers \cite{wilson2011better,ye2024deep,li2026energy}.

This paper introduces nontrivial extensions of the traditional deadline scheduling model to include (i) colocated stochastic supply for a profit-maximizing economic agent in an energy system, and (ii) nonlinear (piecewise linear) pricing of electricity that differentiates the import and export costs. Specifically, the agent faces two types of flexible demands: deferrable demands that arrive randomly and can be delayed within specified deadlines, and nondeferrable price-responsive demands that are always present. The demands are electrical and can be served by grid-imported electricity at the market price or a local stochastic supply with zero marginal cost, such as colocated renewable generation, with the option of exporting local supply to the grid to maximize profit.

The problem of serving flexible demand with colocated stochastic supply arises from the growing need to connect large loads to the power grid. Because of generation and transmission constraints, the grid operator strongly prefers large loads that are flexible and have colocated generation. Deferrable demand arises in many settings: for AI data centers, large language model training and fine-tuning tasks are deferrable with different priority levels defined by completion deadlines \cite{colangelo2026ai}; for large EV-charging service providers, deferrable demand comes from customers who accept relaxed charging deadlines in exchange for reduced prices. Typical examples of nondeferrable price-responsive demands, on the other hand, include heating, ventilation, and air-conditioning (HVAC) systems, data center cooling, server baseline operation, and lighting, whose consumption levels can be modulated in response to electricity prices but cannot be delayed. In all these cases, the economic agent operating such colocated systems aims to maximize economic surplus, responding to the randomness of demand, supply, and electricity prices. 

\subsection{Related works}
We highlight major publications on the joint scheduling of deferrable and nondeferrable flexible demands, focusing on those involving colocated generation resources. At the outset, almost all existing deferrable demand scheduling techniques adopt a wholesale-market pricing model, where electricity prices are linear and stochastic. This paper studies the problem under the regulated retail market pricing model, which is deterministic and piecewise linear with distinct import and export prices. In this setting, the work most relevant to this paper is \cite{alahmed2022net} on the scheduling of nondeferrable, price-responsive demands, which lack inter-temporal dependencies introduced by deferrable jobs. To our best knowledge, this paper is the first on the joint scheduling of deferrable and nondeferrable demands with colocated stochastic supply.

The first work to consider scheduling deferrable demands with colocated stochastic generation appears to be \cite{papavasiliou2010supplying}, where the authors present a finite-horizon Markov Decision Process (MDP) formulation with uncountable state and action spaces (as in our model). An approximate dynamic programming solution is proposed by quantizing the state and action spaces, making the problem tractable but at high computation cost, yielding only a suboptimal solution.

The next group of publications \cite{xu2016dynamic, jin2020joint,jin2020optimal, jia2021structural} share similar MDP models for deferrable jobs and colocated stochastic generation. These models include stochastic wholesale pricing of electricity, colocated random generation, colocated storage, and the ability for deferrable demands to export electricity back to the grid. A key assumption made in these works is that the state and action spaces are finite, making them significantly different from our model in this paper.

Obtaining low-complexity deadline-scheduling algorithms has always been the main focus of job scheduling since the early days of developing the optimal scheduling theory. To this end, one must overcome the curse of dimensionality in the MDP formulation by seeking structural solutions such as priority rules and threshold policies. Most noticeable are the less-laxity and longer-remaining-processing-time (LLLP) \cite{xu2016dynamic} and less-laxity first and later deadline (LLF-LD) priority rules \cite{jin2020joint}. A particularly strong characterization of the optimal policy is obtained in \cite{jin2020optimal,jin2020joint} under a fixed grid-import purchasing policy. In general, these priority rules define only partial priority orderings, which can be used to obtain strong approximate solutions. In addition, threshold and index policies \cite{kim2011scheduling,roozbehani2014robust, yu2018deadline} are also low-complexity solutions. These approaches, however, do not include stochastic colocated resources. Thus, their solution structures are not applicable to the scheduling problem considered here.

Two types of approaches outside the MDP formulation have also been applied to the scheduling problems considered in this paper. One is the receding-horizon model predictive control (MPC) heuristics, which are widely used and serve as standard performance benchmarks \cite{tang2016model, ito2017model, chen2013mpc, yu2013modeling}. Despite lacking firm theoretical justifications, these techniques perform well in practice. The computational cost of these techniques in each decision epoch can be prohibitive. Another group of approaches is based on Lyapunov optimization \cite{neely2010efficient,jin2014optimized,zhou2017optimal}, which accommodates perhaps the most general types of random deferrable jobs and colocated generation. First proposed in \cite{neely2010efficient}, this approach transforms the scheduling of deferrable jobs into a resource allocation problem in a queueing-theoretic framework and applies the back-pressure (Lyapunov drift) scheduling method \cite{tassiulas1992stability,neely2006energy}. These techniques aim to complete all deferrable jobs by stabilizing the job queue. Unfortunately, completion of all jobs by their individual deadlines are not guaranteed. 

When the underlying MDP model is partially or entirely unknown, a natural approach is to apply machine learning techniques. A significant amount of recent literature focuses on applying reinforcement learning (RL) methodologies to schedule deferrable demands, especially for EV charging applications \cite{ye2021model, wu2018optimizing, wan2018model, mocanu2018line, yang2025deep, li2019constrained, neil2010residential}. However, these works do not exploit the structure of the optimal policy.

A closely related line of works \cite{roy2021online, park2023adaptive, nakhleh2022deeptop, hao2022laxity} develops RL algorithms that directly learn the optimal thresholds, leveraging the threshold structure of the underlying policy. In \cite{roy2021online}, a stochastic approximation (SA) algorithm is proposed to learn multiple ordered thresholds. The work in \cite{park2023adaptive} also applies an SA-based approach to learn the optimal threshold policy for an inventory control problem. The authors of \cite{nakhleh2022deeptop} employ a Deep Deterministic Policy Gradient (DDPG)-based approach to learn a binary-action threshold policy. In \cite{hao2022laxity}, a Soft Actor-Critic (SAC)-based threshold learning algorithm for the EV charging station problem is proposed. However, these learning algorithms only consider scheduling of a single type of job. In contrast, the joint scheduling of deferrable and nondeferrable jobs considered here induces a richer optimal policy structure, which calls for a new threshold-learning formulation.

An alternative to the RL approach is the use of evolutionary strategies to obtain low-complexity MDP solutions \cite{salimans2017evolution}. Domain structure can be incorporated into the otherwise black-box evolutionary strategy framework to improve scalability and parallelization. See \eg \cite{bi2023self}. As black-box approaches, however, such algorithms do not account for the inherent policy structure, such as the procrastination structure obtained in this work.

\subsection{Summary of results and contributions}
The objective of this work is to uncover the structural properties of the optimal scheduling policy, from which we derive practical solutions. The main contribution of this work is threefold. First, we extend the classical deadline scheduling problem to (i) enable the joint scheduling of deferrable demands with deadlines and price-responsive nondeferrable demands, (ii) include stochastic colocated supply, (iii) allow deterministic, time-varying, and piecewise-linear pricing of electricity, and (iv) optimize an economic surplus objective.

Second, under an MDP formulation with uncountable state and action spaces, we establish the optimality of a {\em procrastination policy}, which reduces the infinite-dimensional policy space to a low-dimensional Euclidean space parameterized by three procrastination thresholds.

Third, we propose a {\em Procrastination Threshold Reinforcement Learning} (PTRL) algorithm, a structured RL approach that learns the procrastination parameters. We demonstrate using out-of-sample real-world data traces that PTRL outperforms standard reinforcement learning and MPC benchmarks.


\subsection{Mathematical notations}
The notations used in the paper are standard. Vectors are in boldface, with ${\bm x}=(x_1, \ldots, x_N)$ denoting a column vector. We use $\mathbf 1$ for a column of ones with appropriate size, and $\mathds{1}(\cdot)$ as an indicator function that maps to 1 if the logical argument is true and zero otherwise. The notation $[x]^+$ (with square bracket) represents the positive part of $x$.


\section{Problem Formulation} \label{sec:problemformulation}
\begin{figure}[h]
	\centering
	\includegraphics[width =0.9\linewidth]{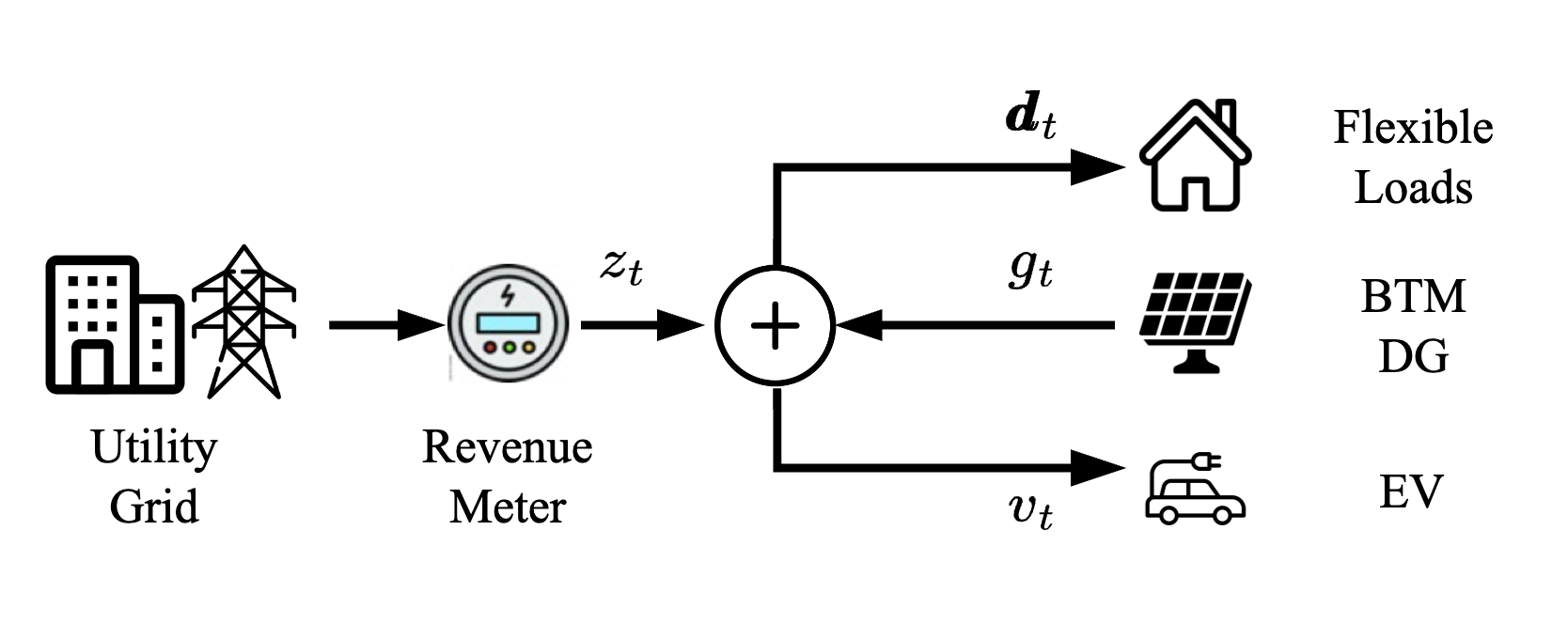}
	\caption{\small A representative joint scheduling model for deferrable and nondeferrable demands, shown here for a household with EV charging (deferrable) and flexible loads (nondeferrable) served by BTM DG. The arrow indicates the direction of power flow when the associated variable is positive.}
	\label{fig:hems-scheme}
\end{figure}

We introduce a discrete-time stochastic dynamic programming formulation for the optimal scheduling of deferrable and nondeferrable demands, using EV charging in a household as a concrete example. As shown in Fig.~\ref{fig:hems-scheme}, the household has a deferrable demand in the form of EV charging served by $v_t$ in interval $t$, a nondeferrable (vector) demand ${\bm d}_t$ (such as heating/cooling, lighting, etc.), and a stochastic behind-the-meter (BTM) distributed generation (DG) $g_t$ that models rooftop solar. The scheduler also has the option to purchase external energy supply $z_t$ from the utility at a cost.

The problem is to schedule the optimal quantity served $(\bm d_t, v_t)$ given the realized random DG $g_t$. For future reference, designated symbols are listed in Table~\ref{tab:my_label}.

\begin{table}[h]
	\captionsetup{font=scriptsize }
	\caption{NOTATIONS FOR MAJOR VARIABLES}
	\vspace{-1ex}
	\centering
	\begin{tabular}{l l}
		\toprule
		Symbol &  Descriptions
		\\   \midrule
		$\bm a_t$ & Action \\
		${\bm d}_t$ & Nondeferrable demand service rate vector  \\
		$\bar {\bm d}$ & Nondeferrable demand maximum service rate \\
		$g_t$ & Distributed generation (DG) \\
		$\bm \mu:=(\mu_t)$ & Scheduling policy \\
		$P_{\pi_t}$ & NEM payment function \\
		$\pibf_t$ & NEM price vector \\
		$\pi_t^+, \, \pi_t^-$ & NEM import and export rates \\		
		$q(\cdot)$ & Non-completion penalty function for deferrable demand \\
		$r_t$ & Reward function \\
		$t,\mathcal T$ & Time index and set\\
		$U_t$ & Nondeferrable demand utility function\\		
		$\partial U_t, \partial U_{it}$ & Marginal utility function\\
		$v_t$ & Deferrable demand service rate\\
		$\bar v$ & Deferrable demand maximum service rate\\
		$\bm x_t$ & State ${\bm x}_t=(y_t,g_t)$ \\
		$y_t$ & Remaining deferrable demand quantity \\
		$z_t$ & Net Consumption of household\\
		\bottomrule
	\end{tabular}
	\label{tab:my_label}
\end{table}

\vspace{-0.5em}

\subsection{The Energy Management and Control}
The household energy consumption is scheduled by an energy management system (EMS) under a $T$-horizon scheduling model indexed by $t=1,\cdots, T$, with
$t=T+1$ the (non-scheduling) terminal interval.

Two types of household energy-consuming devices are considered: one has a deferrable demand (such as EV charging) whose total demand needs to be met by a specific deadline. The other type includes a set of nondeferrable demands that must be served in each interval $t$. At the beginning of interval $t$, the EMS measures the available DG level $g_t$ and sends a control signal $\bm a_t=( \bm d_t, v_t)$ to these devices.

With local generation $g_t$ offsetting the gross consumption, the {\em net consumption} $z_t$ of the household  is given by
\begin{equation}\label{eq:z_t}
z_t=v_t+{\mathbf 1}^\top {\bm d}_t-g_t.
\end{equation}
The EMS's control policy makes the household a {\em prosumer}: a paying {\em consumer} to the distribution utility when $z_t > 0$, a paid {\em producer} by the utility when $z_t < 0$, and a {\em net-zero} customer when $z_t = 0$.

The goal of the EMS is to maximize the household surplus defined by the total benefits minus the payment to the utility. Supply and demand models are detailed next.

\subsection{Supply and costs}  Two power supplies are used to serve both types of the demand.  One is the purchased power from the grid---a decision variable with cost; the other is the BTM DG (rooftop solar), whose production $g_t$ in interval $t$ can be measured at the beginning of interval $t$ but cannot be controlled. The DG production $(g_t)$ is modeled as a non-homogeneous Markovian process with a conditional probability density function $f_t$:
\begin{equation} \label{eq:Markov}
	g_{t+1} \sim f_t(\cdot \, | \, g_t).
\end{equation}
The EMS controls the usage of $g_t$; it may export a fraction of unused $g_t$ to the grid for profit.

The cost of household consumption comes from the power purchased from the grid and the terminal cost of incompletion. We assume that the household is served by a distribution utility under a regulated tariff. A widely adopted tariff is the Net Energy Metering (NEM) that charges or credits the household based on their net consumption $z_t$.

The NEM tariff is deterministic, possibly time varying, and known to all retail customers. In interval $t$, the general form of NEM tariff is defined by a price vector $\pibf_t = (\pi^+_t, \pi^-_t)$.  From the perspective of a customer, $\pi^+_t$ is the import (buy) rate and  $\pi_t^-$ the export (sell) rate. We omit any fixed (connection) charge because it plays no role in optimal scheduling.

The standard NEM payment function $P_{\bm\pi_t}$ is piecewise linear given by
\begin{equation*}
	P_{\pibf_t} (z_t) = (z_t\pi_t^+)\mathds 1(z_t \ge 0)  + (z_t\pi_t^-)\mathds 1(z_t < 0).
\end{equation*}
Note that a negative payment means that the customer is paid by the utility for its net production. The NEM tariff implemented in practice sets the import rate $\pi_t^+$ by the consumer no smaller than the export rate,  $\pi_t^+\ge \pi_t^-$, making the payment function $P_{\pibf_t}$ convex with respect to the net consumption $z_t$.

\subsection{Demand and Benefits of Consumption}
    \subsubsection{Nondeferrable demand}
    The nondeferrable demand is a price-responsive demand that has a consumption bundle ${\bm d}_t := (d_{1t}, \ldots, d_{Kt})$ in interval $t$ from $K$ devices as a decision variable with constraint ${\bm 0}\le \bm d_t \le \bar{\bm d}$, or
    \begin{equation}
        0 \le d_{it} \le \bar d_{i}, \quad \forall i = 1, \ldots, K.
    \end{equation}
    We assume that the household's consumption preference is represented by a differentiable additive time-varying utility function $U_{t}$ with $U_{it}$ being the utility function for the $i$th device in interval $t$:
    \begin{equation*}
        U_t({\bm d}_t) := \sum _{i=1}^K U_{it}(d_{it}).
    \end{equation*}
The marginal utility of the $i$th device in interval $t$ is denoted as $\partial U_{it} := U_{it}'$ and its inverse as $\partial U_{it}^{-1}$. The utility functions have already been learned and known to the EMS scheduler.

\begin{remark}
	{\normalfont Our model can be extended to include stochastic, nondeferrable demand, which is price-insensitive, by defining the DG process $g_t$ as a net renewable variable---the remaining generation after offsetting the inflexible demand---and allowing $g_t$ to have negative values.}
\end{remark}

    \subsubsection{Deferrable demand (EV)}
    The deferrable demand is modeled after the charging of an electric vehicle.  The demand arrives at the beginning of a random interval $t_1$ with a request to consume $y_1$ (kWh) of electricity by the end of interval $t_T$. Without loss of generality, we assume $t_1=1$ and the deadline is the scheduling horizon, $t_T=T$.

    The remaining demand at the beginning of interval $t$ is denoted by $y_t \in [0, y_1]$ and the service rate of the control by $v_t \in [0, \bar v]$.  Assuming unit-length scheduling intervals, the evolution of $y_t$ is
    \begin{equation} \label{eq:yt}
        y_{t+1} = y_t - \eta v_t,
    \end{equation}
    where $\eta$ represents charging efficiency. Without loss of generality\footnote{Remaining demand and charging rate can be scaled by $1/\eta$.}, we assume that $\eta = 1$.  With the maximum charging rate $\bar{v}$,  the amount of charging in a single interval $v_t$ satisfies
    \begin{equation}
    	v_t \in [0, \min\{y_t, \bar v\}].
    \end{equation}

    We allow charging demand not met at the end of scheduling session  and impose a penalty  $q(y_{T+1})$  on the unsatisfied demand $y_{T+1}$. The penalty function $q$ is assumed to be strictly increasing and convex.

\subsection{The Stochastic Dynamic Optimization}
    We formulate the optimal energy management problem as a finite-horizon, discrete-time, and continuous-state constrained MDP $\mathcal M = (\mathcal S, \mathcal A, r, T, \mathbbm P)$, with scheduling horizon $\cT= \{1, \cdots, T\}$, state vector ${\bm x}_t = (y_t, g_t) \in \mathcal S \subseteq \mathbb R_+^2$, and action vector ${\bm a}_t = (\bm d_t, v_t) \in \mathcal A \subseteq \mathbb R_+^{K+1}$. The stage reward is the surplus of the household:

    \begin{equation}
        r_t(\bm x_t, {\bm a}_t) = U_t(\bm d_t) - P_{\pibf_t}(v_t + \mathbf{1}^T \pmb d_t - g_t).
    \end{equation}
The terminal reward is the penalty on the unsatisfied deferrable demand at its deadline: $r_{T+1}(\bm x_{T+1}) = -q(y_{T+1})$.
	
    We formulate a stochastic dynamic optimization problem over a sequence of policies $\bm \mu = (\mu_1, \ldots, \mu_T)$, where,  in each interval,  $\mu_t: \bm x_t \mapsto {\bm a}_t=({\bm d}_t, v_t)$, and the expected total reward is maximized
    \begin{align} \label{eq:opt0}
    \begin{array}{rl}
        \underset{\bm \mu=(\mu_1,\cdots,\mu_T)}{\mbox{maximize}} \quad  &\mathbbm E \left[\sum_{t = 1}^T  r_t\big ( \pmb x_t, \mu_t (\pmb x_t) \big) - q(y_{T+1}) \right] \\[0.7em]
        \text{subject to} & (1) - (5).
        \end{array}
    \end{align}

Let ${\bm \mu}^* =(\mu_1^*,\cdots, \mu_T^*)$ be the optimal policy  in (\ref{eq:opt0}).  Define the optimal reward-to-go function $V_t$ for $t=1,\cdots, T$ and for all $\bm x \in \mathcal S$,
\[
V_t(\bm x):=\mathbb E\Bigg[\sum_{k=t}^T r_k(\bm x_k,\mu_k^*(\bm x_k)) - q(y_{T+1}) \; \bigg | \; \bm x_t=\bm x\Bigg].
\]
By defining the terminal value function as $V_{T+1}(\pmb x_{T+1}) = -q(y_{T+1})$, the Bellman principle of optimality requires, for all $t = 1, \ldots, T$ and $\bm x$,
    \begin{equation} \label{eq:bellmanequation}
        V_t(\bm x) = \max_{\bm a \in \mathcal A(\bm x)} \; \Big(r_t(\bm x, \bm a) + \mathbb E[V_{t+1}(\bm x_{t+1})\mid \bm x_t=\bm x]\Big).
    \end{equation}
\subsection{Model Assumptions}
We summarize key model assumptions with comments and justifications.
    \begin{enumerate}[label=A\arabic*)]
    \item The utility function of nondeferrable consumption is differentiable, concave, increasing, and known.
    \item The set of NEM price vectors $\{\pibf_{t}=(\pi_t^+,\pi_t^-)\}$ are deterministic and known to the EMS, satisfying
        \begin{equation} \label{eq:pi}
        \big(\pi_t^+\ge \pi_t^-, \forall t\big)~\mbox{and}~\big(\min\{\pi_t^+\}\ge \max\{\pi_t^-\}\big).
        \end{equation}
    \item The penalty function $q$ for incomplete deferrable demand  is strictly increasing and convex with  $\frac{dq}{dy}(0) > \max\{\pi _t^+\}$.
    \item Without loss of generality, $y_1 \le T\bar{v}$.
    \end{enumerate}
A1 is standard in microeconomic theory of consumer behavior. The exact utility function of nondeferrable demand may be unknown in practice, but it can be recovered from historical price-quantity data using established techniques, such as Afriat's Theorem \cite{afriat1967construction} for nonparametric estimation or inverse optimization \cite{keshavarz2011imputing} for parametric models. Assumption A2 on deterministic NEM prices is realistic, as NEM rates are set by regulated distribution utilities and are publicly available. Relaxing this assumption to stochastic prices corresponds to the wholesale market setting, where the payment function is linear rather than piecewise linear as in NEM. This setting is not the focus of the present study. The relation between $\pi_t^-$ and $\pi_t^+$ rules out risk-free arbitrage, which holds for all practical NEM tariffs.

A3 eliminates the trivial case that it is more economic not to serve the deferrable demand than to serve it with purchased power.The convexity of the penalty function is a design choice to minimize the incomplete demand. This is a standard modeling approach in the deadline scheduling literature \cite{roozbehani2014robust, kim2011scheduling, jin2020optimal, zhou2017optimal, xu2016dynamic, jia2021structural, jin2020joint, yu2018deadline}. If A4 is not satisfied, there will be at least  $q(y_1-T\bar{v})$ penalty regardless of the scheduling algorithm. Therefore, we only need to consider serving deferrable demand of size  $y_1\le T\bar{v}$. By assumption A4, the completion of charging demand is always feasible.

\section{Structures of the Optimal Policy}\label{sec:thresholdpolicy}
In this section, we introduce the procrastination policy and establish its optimality.
\subsection{Structure of optimal scheduling with deferrable demand} \label{subsec:procras}
We motivate the {\em procrastination policy} by first considering the simple scenario that schedules only the deferrable demand (EV charging) and the NEM tariff rates are time invariant, \ie  $\pibf_t = (\pi^+, \pi^-)$ for all $t$.  Because the marginal cost of DG is zero, it is intuitive to avoid purchasing power (at the price of $\pi^+$) by delaying (procrastinating) any purchase unless it is necessary. The timing of such purchasing decisions needs to be made precise.

The procrastination policy can be stated as follows:  in interval $t$ with remaining demand of $y_t$,
    \begin{enumerate}
    	\item \textit{use all DG production $g_t$ to meet the demand and export surplus.}
    	\item \textit{delay (procrastinate) purchasing until it is no longer possible to fulfill the deferrable demand by the end of interval $T$.}
    \end{enumerate}
The intuition of procrastination policy is as follows.
Item 1) prioritizes using DG $g_t$ to serve the demand because DG has zero marginal cost. Serving demand with $g_t$ has a higher marginal value ($\pi^+$) than that gained from exporting $g_t$ to the grid. Item (2) means that there is no purchase of power from the grid if the remaining demand can still be served by the deadline.

We establish formally (Appendix \ref{sec:appendixpropproof}) the optimality and the threshold structure of the procrastination policy $(\mu_t^{\text{p}})$.

\begin{proposition}[Optimality of Procrastination]\label{prop:1}
Under the time-invariant NEM pricing and in the absence of nondeferrable demand, the procrastination policy $\mu_t^{\text{\normalfont p}}:\bm x_t \mapsto v_t^*$, with procrastination threshold $\theta_t:=(T-t)\bar{v}$ is optimal :
\begin{align}
v^*_t & =		\begin{cases}
y_t, & 0 < y_t \le  \min\{\bar v, \; g_t\}\\
\min\{\bar v, \; g_t\},  & \min\{\bar v, \; g_t\} <  y_t \le  \theta_t + \min\{\bar{v},g_t\}\\
\min\{y_t - \theta_t, \bar v\} &  \theta_t + \min\{\bar v, \; g_t\} < y_t.
  			\end{cases}\label{eq:procras_chargingrate}
  		\end{align}
\end{proposition}

\begin{figure}[t]
\hspace{-1em}
    \includegraphics[width=\columnwidth]{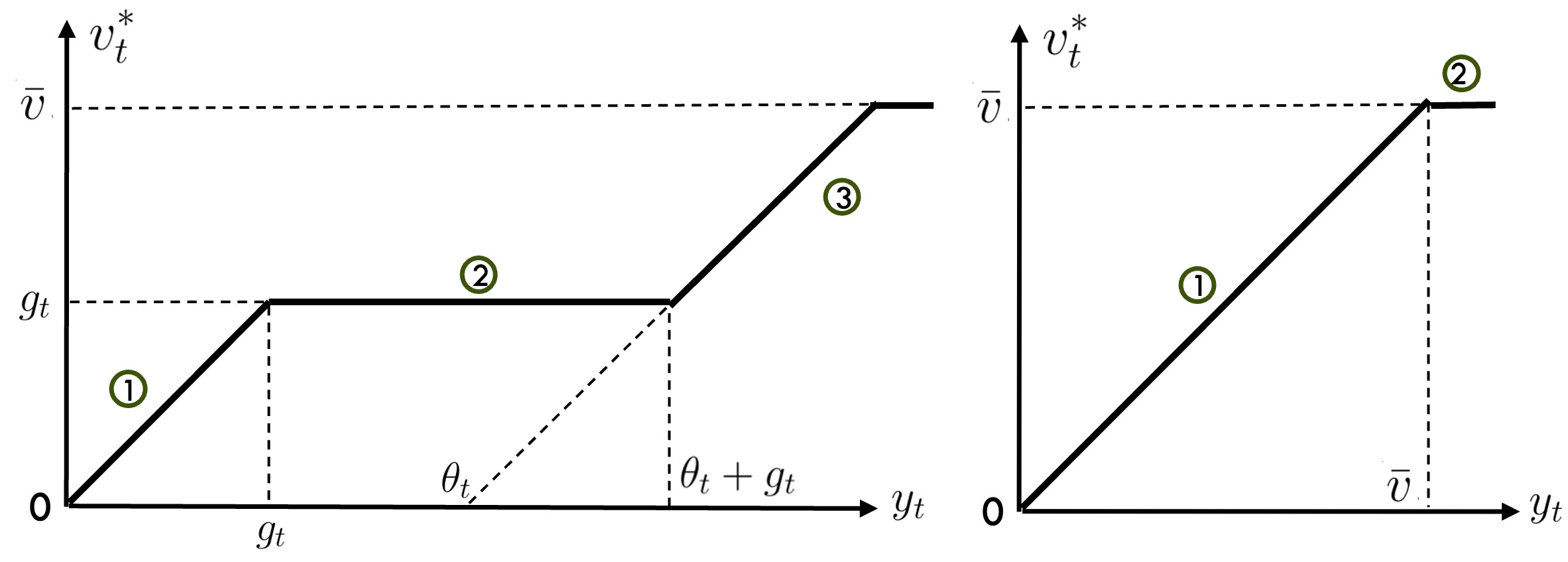}
    \caption{\small Procrastination scheduling with deferrable demand under time-invariant NEM. Left: DG level $g_t \le \bar{v}$. Right: DG level $g_t > \bar{v}$.}
    \label{fig:Procst1}
\end{figure}

Fig.~\ref{fig:Procst1} illustrates the optimal policy with the left figure for the  case $g_t\le \bar{v}$. Segment $\textcircled{{\scriptsize 1}}$ is when the remaining demand is small, $y_t < g_t$,  then $y_t$ is met with $g_t$.  When the remaining demand $y_t$ exceeds the local generation $g_t$ but the deadline is afar (segment $\textcircled{{\scriptsize 2}}$), it is optimal to procrastinate to purchase power, serving the demand using all local generation. When the remaining demand is high and deadline is near, the incompletion penalty is unavoidable unless it is reduced with purchased power as shown in segment $\textcircled{{\scriptsize 3}}$.

Segment $\textcircled{{\scriptsize 3}}$ then gives an interpretation of $\theta_t$ as {\em the laxity of the deferrable demand}, \ie $\theta_t$ is the maximum remaining demand level that, with DG $g_t$, the incomplete penalty at the end of scheduling horizon can be avoided. Therefore, if $y_t>\theta_t+g_t$, procrastination must end.

Note that the optimality of the procrastination policy holds even when the local generation $(g_t)$ is nonstationary and arbitrarily correlated.   Unfortunately,  policy $\mu_{\mathrm{p}}$ defined in (\ref{eq:procras_chargingrate}) is no longer optimal when the NEM tariff is time varying.  Consider the case when we have the time-of-use (ToU) NEM defined by the off-peak period price $(\pi^+_{\mbox{\scriptsize off}},\pi^-_{\mbox{\scriptsize  off}})$ and the on-peak period price $(\pi^+_{\mbox{\scriptsize  on}},\pi^-_{\mbox{\scriptsize  on}})$,  that satisfy $\pi^{+}_{\mbox{\scriptsize  on}} > \pi^+_{\mbox{\scriptsize off}} > \pi^-_{\mbox{\scriptsize on}} > \pi^{-}_{\mbox{\scriptsize  off}}$. 

Suppose that the deferrable demand starts during the low-price off-peak period when the purchasing price $\pi_{\text{off}}^+$ is low, and the demand cannot be met within the off-period; purchasing power from the grid during the on-peak periods with higher price $\pi_{{\text{on}}}^+$ is necessary with high probability. Then procrastinating purchasing during the off-peak period can be suboptimal. Whether to procrastinate or not depends on the amount of DG available in the future, and the procrastination threshold should be set by the risk of having to import power during on-peak periods.  

We now consider the optimal scheduling of deferrable demand under the general time-varying NEM tariff.  We show that the optimal scheduling policy has the same structure of the procrastination policy given in (\ref{eq:procras_chargingrate}) except that the thresholds depend on the available DG level $g_t$.

\begin{figure}[t]
	\centering
	\includegraphics[scale=0.57]{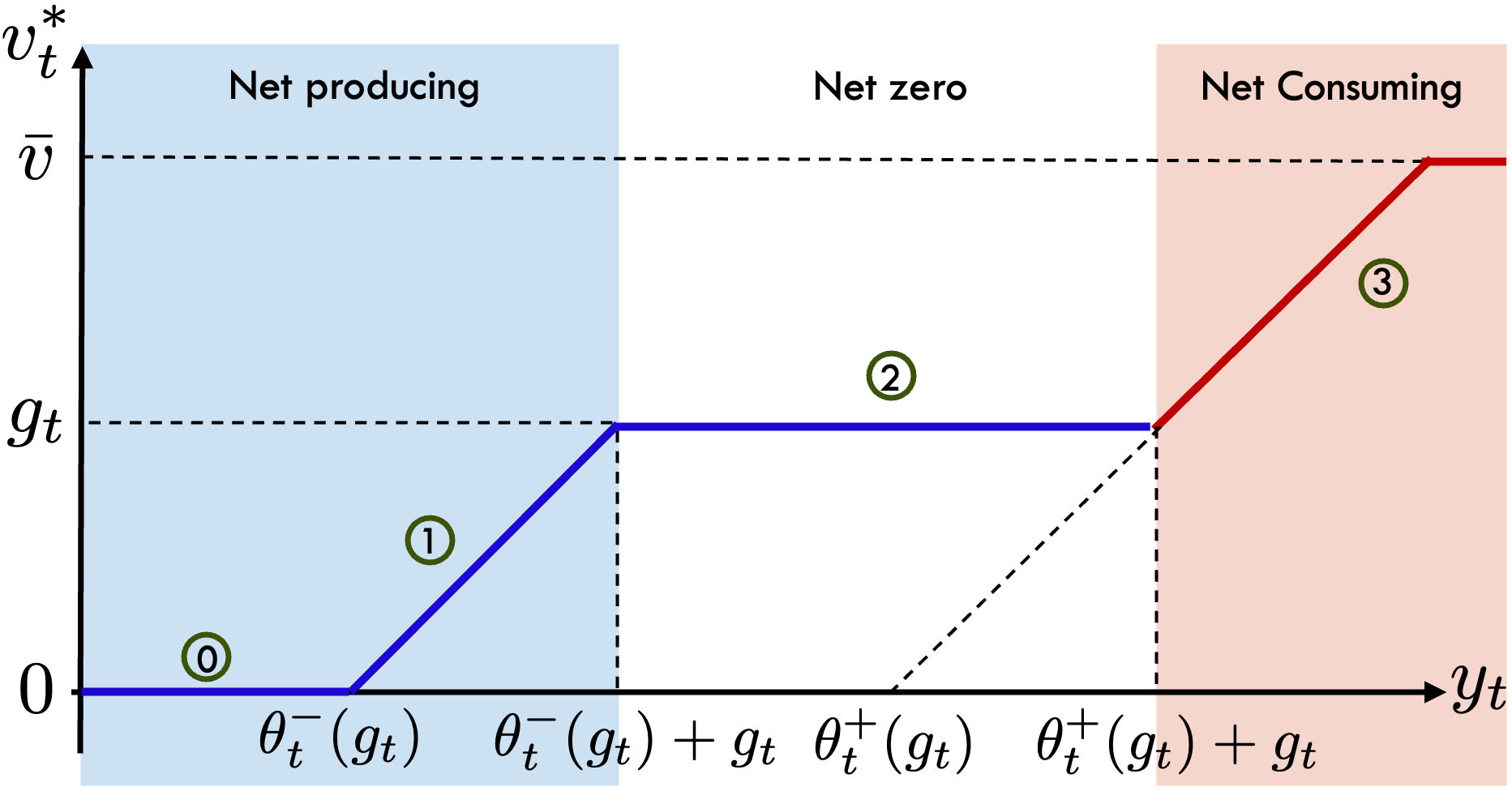}
	\caption{\small Procrastination scheduling of deferrable demand with Markovian DG $g_t < \bar{v}$.}
	\label{fig:Procst2}
\end{figure}

\begin{theorem}[Threshold structure of optimal deferrable demand scheduling] \label{thm:opt1}
Assume that the local generation $(g_t)$ is non-homogeneous Markovian.
In the absence of nondeferrable demand, the optimal scheduling policy $\mu_t^*: \bm x_t \mapsto v_t^*$
 is defined by two procrastination thresholds on the remaining demand $y_t$, $0 \le \theta_t^- (g_t) \le \theta_t^+ (g_t) \le (T-t)\bar{v} $:
	\begin{equation} \label{eq:vt}
v^*_t=\begin{cases}
0, & \text{if}~y_t  \le \theta_t^-(g_t)\\
y_t-\theta_t^-(g_t), & \mbox{if $\big(y_t > \theta_t^-(g_t)\big)$ and}\\
&\mbox{$\big(y_t \le \theta_t^-(g_t)+\min\{\bar{v}, g_t\}\big)$}\\
\min\{\bar{v}, g_t\}, & \text{if}~ \big(y_t > \theta_t^-(g_t)+\min\{\bar{v}, g_t\}\big)\\
 & \text{and}~\big(y_t \le \theta_t^+(g_t)+\min\{\bar v, g_t \}\big)\\
\min\{y_t - \theta_t^+(g_t),\bar{v}\}, &   \text{otherwise,} \\
			\end{cases}
		\end{equation}

		where $\theta_t^+(g_t)$ and $\theta_t^-(g_t)$ are defined by the following first-order optimality conditions on (\ref{eq:bellmanequation}):
		\begin{align*}
			\partial_y^+ \bar V_{t+1}(\theta_t^+(g_t); g_t) \le -\pi_t^+ \le \partial_y^- \bar V_{t+1}(\theta_t^+(g_t); g_t), \\
			\partial_y^+ \bar V_{t+1}(\theta_t^-(g_t); g_t) \le -\pi_t^- \le \partial_y^- \bar V_{t+1}(\theta_t^-(g_t); g_t).
		\end{align*}
		Here, $\bar V_{t+1} (y,g):= \mathbb E_{g_{t+1}}[V_{t+1}(y, g_{t+1})\,|\,g_t = g]$, and $\partial_y^+, \partial_y^-$ denote its right and left derivatives with respect to $y$.
\end{theorem}

The similarity of (\ref{eq:vt}) to the time-invariant NEM case (\ref{eq:procras_chargingrate}) is evident. Specifically, setting $\theta_t^-(g_t) = 0$ and  $\theta_t^+(g_t) = (T-t)\bar v$ in (\ref{eq:vt}) recovers the time-invariant NEM case (\ref{eq:procras_chargingrate}), confirming that Proposition~\ref{prop:1} is a special case of Theorem~\ref{thm:opt1}. The scheduling figure shown in Fig.~\ref{fig:Procst2} for the case $g_t<\bar{v}$ is the same as the invariant NEM case in Fig.~\ref{fig:Procst1} (left), except that the threshold $\theta_t$ in Fig.~\ref{fig:Procst1} is replaced by  $\theta_t^+(g_t)$ and the $y$-axis of Fig.~\ref{fig:Procst1} is moved left by $\theta_t^-(g_t)$ with a new segment $\textcircled{{\scriptsize 0}}$,  between $y_t=0$ and $y_t=\theta_t^-(g_t)$ where $v_t^*=0$.

The presence of segment $\textcircled{\scriptsize 0}$ with the  new threshold of $\theta^-_t(g_t)$ is intriguing. It implies that, when the remaining demand $y_t$ is small enough (or the deadline of completion is far), $y_t$ is not served by the DG and all $g_t$ is exported to the grid. In other words, procrastination not only applies to purchased power, it also applies to DG.  

Not serving $y_t$ with $g_t$ is counterintuitive because the exporting price $\pi_t^-$ is lower than the minimum of the purchasing price $\min\{\pi_t^+\}$. If not serving $y_t$ requires purchasing in the future, then not serving $y_t$ must be suboptimal.  
However, if there is a high conditional probability that $g_{t+1}>y_t$ and the exporting price $\pi^-_{t+1}<\pi_t^-$, then it may be better to take the chance of exporting $g_t$ and hoping $g_{t+1}> y_t$.
Consider the extreme case of $\Pr(g_{t+1}>y_t\,|\,g_t)=1-\epsilon$ and $\pi^-_{t+1}=\epsilon$. As $\epsilon\rightarrow 0$, the profit from procrastinating serving $y_t$ to the next interval converges to $\pi^-_ty_t>0$ while the profit of serving  $y_t$ immediately converges to zero.

It turns out that the threshold $\theta^-_t(g_t)$ plays the critical role of ensuring the cost of delaying (procrastinating) serving the deferrable demand $y_t$ in segment $\textcircled{\scriptsize 0}$ is justified. More precisely, $\theta_t^-(g_t)$ is designed to hedge against the risk of low profit when there is a chance of high DG level and low exporting price.

Note that, for fixed local generation $g_t$, the optimal scheduling policy has three regions:   a net-producing region when the remaining demand is small, a net-consuming region when the remaining demand is high, and the interesting  positive interval $[\theta_t^-(g_t)+g_t,\theta_t^+(g_t)+g_t]$ when the household is a net-zero customer.  We will see that this particular feature prevails for the optimal scheduling policy as shown in the next section and Fig.~\ref{fig:priority_rule}.

\subsection{Optimal Scheduling of Nondeferrable Demand} \label{sec:nondeferrableschedule}
For the pure deferrable demand,  Theorem~\ref{thm:opt1} shows that the optimal policy is a two-threshold policy on the remaining demand. We now consider pure nondeferrable demands and show that a two-threshold policy on DG is optimal.

The nondeferrable demand scheduling is  (\ref{eq:opt0}) with $y_1=0$,
which removes the temporal coupling constraint (\ref{eq:yt}), reducing the multi-stage optimization to a sequence of interval-by-interval optimizations:
\[
\begin{array}{rl}
\underset{ \{\bm d_t: d_{it} \in [0,\bar{d}_i] \}}{\mbox{maximize}} & \Big(\sum_{i=1}^K U_{it}(d_{it}) - P_{\bm \pi_t}(z_t)\Big)\\
\mbox{subject to} & z_t={\bm 1}^\top \bm d_t-g_t.
\end{array}
\]
To understand the structure of the above optimization, we ignore the constraints $0\le d_{it}\le \bar{d}_i$.
Suppose that $g_t$ is small enough that the household is a consumer, \ie $z_t=\sum_i d_{it}-g_t\ge 0$. Then $P_{\bm \pi_t}(z_t)=\pi^+_t z_t$.  The first order optimality condition gives
\[
\partial{U}_{it}(d^*_{it}) = \pi_t^+~~\Rightarrow~~d^*_{it} =\partial{U}^{-1}_{it}(\pi_t^+).
\]
Let $d_{it}^+:=\partial{U}^{-1}_{it}(\pi_t^+)$.
We have the optimal consumption $d_{it}^*=d_{it}^+$ as a consumer, for all $g_t \le d^+_{t}:=\sum_i d^+_{it}$.

\begin{figure}[h]
    \centering
    \includegraphics[scale=0.55]{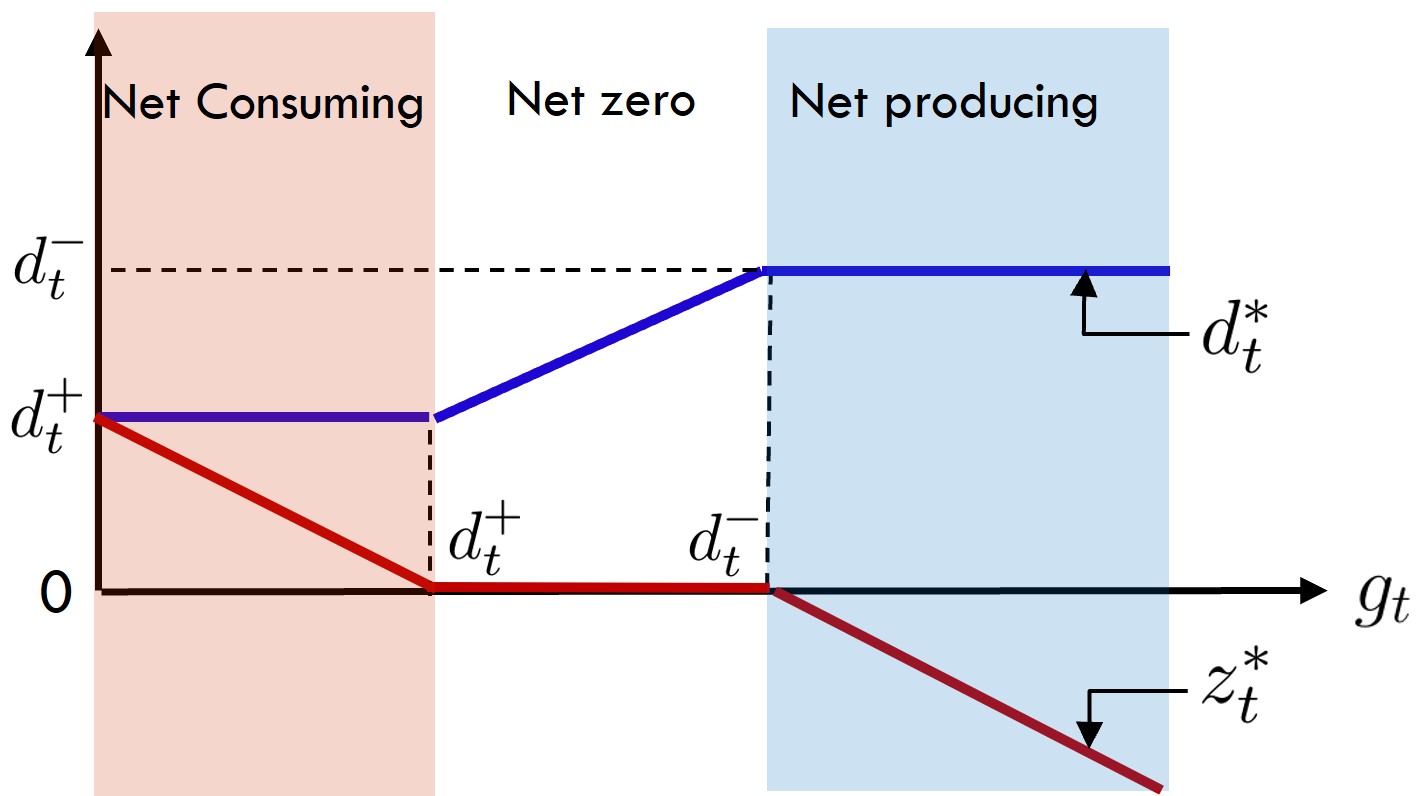}
    \caption{\small Net-consumption zones, the optimal total demand $d^*_t:=\mathbf{1}^\top {\bm d}_t$, and the (optimal) net consumption; $z_t^*:=d^*_t-g_t$.}
    \label{fig:zone}
\end{figure}

Following the same argument by considering the household is a producer ($z_t \le 0$) and a net-zero customer ($z_t = 0$) we obtain the optimal consumption policy is a threshold policy on $g_t$:
\begin{equation}\label{eq:dstar}
d_{it}^* = \left\{\begin{array}{ll}
d^+_{it}:= \partial{U}^{-1}_{it}(\pi_t^+), & \mbox{if $g_t \le d^+_{t}:=\sum_i d^+_{it}$}\\
d^-_{it}:=  \partial{U}^{-1}_{it}(\pi_t^-), & \mbox{if $g_t \ge d^-_{t}:=\sum_i d^-_{it}$}\\
d_{it}^*:=  \partial{U}^{-1}_{it}(\pi_t^o), & \mbox{otherwise,}
\end{array}\right.
\end{equation}
where $\pi_t^o$, a function of $g_t$, is such that $\sum_i d_{it}^* = g_t$.  As shown in Fig.~\ref{fig:zone}, the household is a net-consumer with a constant consumption $\bm d_t^*=\bm d_t^+$ when $g_t < d^+_t$, a net-producer with a constant consumption $\bm d_t^*=\bm d_t^-$ when $g_t > d^-_t$, and a net-zero customer when $g_t \in [d^+_t,d^-_t]$. See \cite{alahmed2022net} for the more general cases.

\subsection{Structure of the Optimal Scheduling Policy} \label{subsec:jointscheduling}	
Our main result given in  Theorem~\ref{thm:opt2} below and proved in Appendix \ref{sec:appendixproof} is that the optimal scheduling policy is defined by three (procrastination) parameters ${\bm \theta}_t:=(\theta_t^-, \theta_t^0, \theta^+_t)$, all functions of the state ${\bm x}_t=(y_t,g_t)$.  These parameters define two structural characteristics of the optimal policy: threshold structure on the state space and priority structure of the scheduling. We use the Fig 5, a partition of state space into different net-consumption zones under time independent $g_t$ model, to give a graphical explanation of threshold structure and scheduling priority of the optimal policy, followed by Theorem~\ref{thm:opt2} that gives the formal characterization.
    
\subsubsection{Threshold structure of the optimal policy}
\begin{figure}[t]
	\centering
	\includegraphics[scale = 0.45]{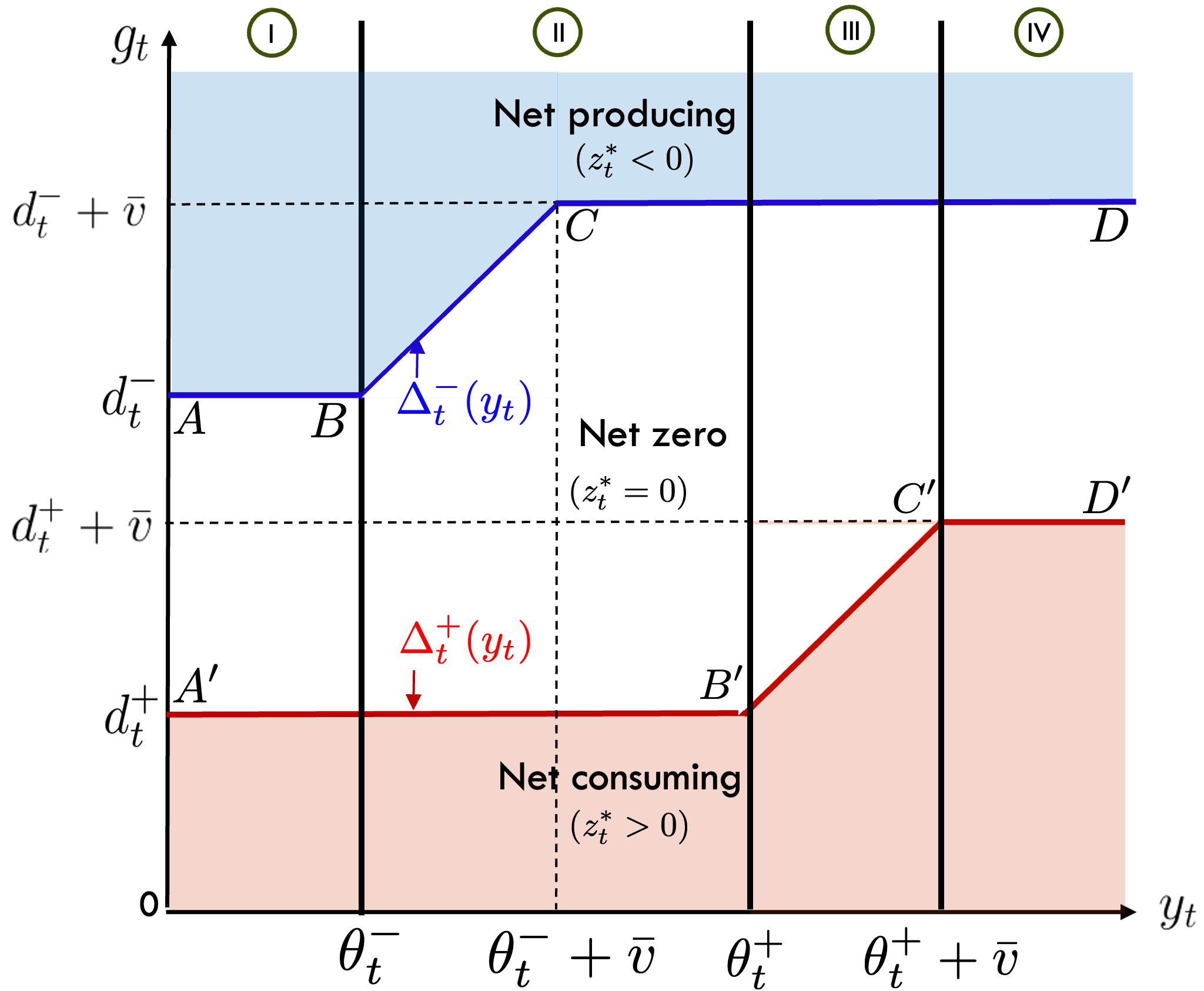}
	\caption{Threshold and priority structures of the optimal policy.}
	\label{fig:priority_rule}
\end{figure}

As part of the optimal policy ${\bm \mu}^*=(\mu_t^*)$,  $\mu^*_t$ is defined on the state space ${\mathscr X}_t$ of ${\bm x}_t$. Fig.~\ref{fig:priority_rule} shows that $\mu^*_t$  partitions ${\mathscr X}_t$ into three net-consumption regions as in the pure nondeferrable demand case: the net-producing region when $z^*_t<0$, the net-consuming region when $z^*_t>0$, and the net-zero region when $z^*_t=0$.  The boundaries of the three regions are characterized by piecewise linear functions $\Delta_t^-$ and $\Delta_t^+$ of the remaining demand $y_t$. Their intercepts ($A$ and $A'$), the break points ($B, C, B',C'$), and asymptotes $(D,D')$ are all set by parameters $(\theta_t^-,\theta_t^+)$ with $d_t^{\pm}$ and $\bar{v}$ being constants.

As in the pure deferrable demand case in Fig.~\ref{fig:Procst2},  $(\theta_t^-,\theta_t^+)$ are threshold parameters on the remaining demand.  Similar to Fig.~\ref{fig:zone},  $(d_t^-,d_t^+)$ are  threshold parameters  on DG level $g_t$, also functions of $(\theta_t^-,\theta_t^+)$. See Theorem~\ref{thm:opt2}.

The partition of the state space into three net-consumption regions significantly reduces the complexity of the optimization inherent in the payment function $P_{\bm \pi_t}$.   It also suggests that the optimal scheduling has a certain priority structure.  The threshold parameters $(\theta_t^-,\theta_t^+)$ represent the critical level of laxity for the remaining demand $y_t$ that requires a change of priority. 

\subsubsection{Priority structure of the optimal policy}

We define a partial priority order $\succeq$ in using the available DG $g_t$ among three actions: serving the nondeferrable demand, serving the deferrable demand, and exporting to the grid. Specifically, action $X$ has priority higher than or equal to  ($\succeq$) action $Y$ implies that the available DG $g_t$ will be devoted to  meet action $X$'s need. Only when $g_t$ is large enough will the surplus DG be used for action $Y$.  The notation  $X \approx Y$ means that  $X$ prioritized over $Y$ is the same as $Y$ over $X$.

The procrastination parameters $(\theta_t^-,\theta_t^+)$ define four segments, $\textcircled{\scriptsize I}$-$\textcircled{\scriptsize IV}$, on the remaining demand axis $y_t$.  Within each segment, the optimal scheduling policy has a priority that defines the optimal action $({\bm d}_t^*, v^*_t)$.
\begin{enumerate}
\item[$\textcircled{\scriptsize I}$] Nondeferrable $\succeq$ Exporting DG $\succeq$ Deferrable.\\
    In this region, $y_t<\theta^-_t$, the remaining deferrable demand is low and the deadline afar.  For  the same reason as in segment $\textcircled{\scriptsize 0}$ in Fig.~\ref{fig:Procst2}, exporting DG is preferred over serving the deferrable. 
    
    The optimal choice between serving nondeferrable demand and exporting renewable is already settled in the pure nondeferrable case: serving nondeferrable has the higher priority over exporting renewable. Therefore, the available DG $g_t$ is used to serve the nondeferrable to its optimal value before exporting $[g_t-d_t^-]^+$ to the grid. All purchased power is used for meeting the nondeferrable demand. At the lowest priority, the deferrable demand is not served.
\item[$\textcircled{\scriptsize II}$] Nondeferrable $\succeq$ Deferrable $\succeq$ Exporting DG.\\
    As in the pure deferrable demand case, when $y_t > \theta^-_t$, serving the deferrable demand has the higher priority over exporting $g_t$ to the grid. However, delay serving the deferrable would not lead to incompletion penalty, making serving the deferrable demand a lower priority than serving the nondeferrable. We thus use all available $g_t$ first to meet the nondeferrable demand.  Only when $g_t>d^-_t$ will the deferrable demand be served by the excess DG $[g_t-d^-_t]^+$, which explains the segment $\overline{BC}$ on the upper boundary $\Delta^-(y_t)$ of the net-zero zone.  Note also that the power purchased in the net-consuming zone are all for the nondeferrable demand, which explains the lower boundary of the net-zero zone.
\item[$\textcircled{\scriptsize III}$] Nondeferrable $\approx$ Deferrable $\succeq$ Exporting DG.\\
    In this region, the remaining demand exceeds the laxity limits. Procrastination ends and the remaining demand must be reduced up to $y_t - \theta_t^+ $ by either DG or purchased power to avoid non-completion penalty. On the other hand, the total demanded quantity is $d_t^+$ from the nondeferrable demand as long as the available DG is below $d_t^+$.  Therefore, the total minimum amount of purchased power is given by the segment $\overline{B'C'}$ on the lower boundary $\Delta^+(y_t)$ of the net-zero zone. As long as the DG level is below $d_t^-+\bar{v}$, there will be no export, which explains the upper boundary of the net-zero region.
\item[$\textcircled{\scriptsize IV}$] Deferrable $\succeq$ Nondeferrable $\succeq$ Exporting DG.\\
 This is the case when the remaining demand is large. It is optimal to reduce $y_t$ maximally, using all available DG and importing power if necessary. Thus the nondeferrable is prioritized over the deferrable over exporting.  Note that the gap between the upper and lower boundaries of the net-zero zone is exactly the width of the net-zero zone in Fig.~\ref{fig:zone}.
\end{enumerate}
\subsubsection{The main theorem}
The above graphical illustration leads to Theorem~\ref{thm:opt2} that completely characterizes the optimal scheduling. See Appendix~\ref{sec:appendixproof} for a formal proof. 

\begin{theorem}[Structure of the Optimal Scheduling Policy]\label{thm:opt2}
        Assume that the DG process $(g_t)$ is Markovian. The optimal scheduling policy $\mu^*_t$ in interval $t$ is
   defined by three parameters $\theta_t^-(g_t) \le \theta_t^0(g_t) \le \theta_t^+(g_t)$. Parameters $\theta_t^-(g_t)$ and $\theta_t^+(g_t)$ are procrastination thresholds on the remaining demand and a pair of net-consumption thresholds  $\Delta_t^+(g_t,y_t)$ and $\Delta_t^-(g_t,y_t)$ (computable from $\theta_t^-(g_t)$ and $\theta_t^+(g_t)$) on the DG $g_t$ that put the household in the {\em net-consuming,} {\em net-producing,} and  {\em net-zero} modes:
        \begin{enumerate}
            \item {\bf Net-consuming:} When $g_t < \Delta_t^+(g_t,y_t)$, where
            \begin{align*}
            \Delta_t^+(g_t,y_t) &:= v^+_t(g_t,y_t)+\sum_{i=1}^K d_{it}^+\\
            v^+_t(g_t,y_t) &:= \min \Big \{ \bar v, [y_t - \theta_t^+(g_t)]^+ \Big \}\\
            d^+_{it} &:= \min \Big \{ \bar d_i, \partial U_{it}^{-1}(\pi_t^+)\Big \},
            \end{align*}
            the household is a {\em net consumer} with net consumption $z^*_t>0$ and the optimal action $(\bm d^*_t,v^*_t)$ given by
            \begin{align*}
                {\bm d}^*_t =(d_{it}^+),~~v_t^* &= v_t^+(g_t,y_t).
            \end{align*}

            \item {\bf Net-producing:}  When $g_t > \Delta_t^-(g_t, y_t)$, where
            \begin{align*}
            \Delta_t^-(g_t,y_t)&:= v^-(g_t, y_t)+ \sum_{i=1}^K d_{it}^- \\
            v_t^-(g_t,y_t) &:=  \min \Big\{ \bar v, [y_t - \theta_t^-(g_t)]^+ \Big\}\\
                d_{it}^- &:= \min\Big\{ \bar d_i, \partial U_{it}^{-1}(\pi_t^-) \Big\},
            \end{align*}
    the household is a {\em net producer} with net consumption $z^*_t<0$ and the optimal action $(\bm d^*_t,v^*_t)$ given by
            \begin{align*}
                {\bm d}^*_t = (d_{it}^-),~~v^*_t=v_t^-(g_t,y_t).
            \end{align*}
            \item {\bf Net-zero:}  When $g_t \in [\Delta_t^+(g_t, y_t), \Delta_t^-(g_t, y_t)]$, the household is a {\em net-zero customer} with net consumption $z^*_t=0$ and the optimal action $(\bm d^*_t,v^*_t) = \mu^*_t(y_t,g_t)$ is given by
            \begin{align}
            & ~v_t^*~~~ = \min \{ \bar v, \big[y_t - \theta_t^0 (g_t)\big]^+\} \label{eq:nz_charging}\\[0.5em]
            &
            \begin{array}{lrl}
            {\bm d}_t^* &= \underset{\{d_{it} \in [0,\bar{d}_i]\}}{\arg\max}  & \sum_{i=1}^K U_{it}(d_{it})\\[1em]
              &\text{\em subject to} & \sum_{i} d_{it} +v^*_t = g_t.
            \end{array} \label{eq:nz_consumption}
            \end{align}
\end{enumerate}
\end{theorem}

Note that Fig.~\ref{fig:priority_rule} is no longer accurate for the Markovian case because the procrastination parameters $(\theta_t^+,\theta_t^0,\theta_t^-)$ all depend on $g_t$.  The above theorem does include the independent DG as a special case, where the procrastination thresholds are independent of $g_t$ and the boundary lines are all vertical in Fig.~\ref{fig:priority_rule}.

The three procrastination parameters are formally defined by the first-order optimality conditions on (\ref{eq:bellmanequation}), with detailed derivations provided in Appendix~\ref{sec:appendixproof}. In general, closed-form expressions of these parameters are unavailable, which motivates the reinforcement learning approach that exploits the low-dimensional threshold structure of the optimal policy.
\section{Procrastination Threshold Reinforcement Learning (PTRL)}\label{sec:thresholdlearning}
	In this section, we propose a \textit{Procrastination Threshold Reinforcement Learning} (PTRL) algorithm that learns the procrastination parameters ${\bm \theta}=({\bm \theta}_1,\cdots, {\bm \theta}_T)$, where ${\bm \theta}_t=(\theta_t^+,\theta_t^0,\theta_t^-)$ is the parameter that defines optimal policy $\mu_t^*$  in Theorem~\ref{thm:opt2}. PTRL is a structured RL algorithm based on SAC \cite{haarnoja2018softarxiv} that reduces $(K+1)$-dimensional continuous action space to a 3-dimensional parameter space. By dimension reduction, PTRL mitigates the high sample complexity from which general RL methods suffer in high-dimensional action spaces \cite{sutton2018reinforcement} and yields stronger generalization guarantees for the learned policy. Throughout this section, for the notational convenience, we augment time index $t$ into the state and denote it as $\hat {\bm x}_t := (\bm x_t, t )$.
	
\begin{figure}[t]
		\centering
		\includegraphics[width = \linewidth]{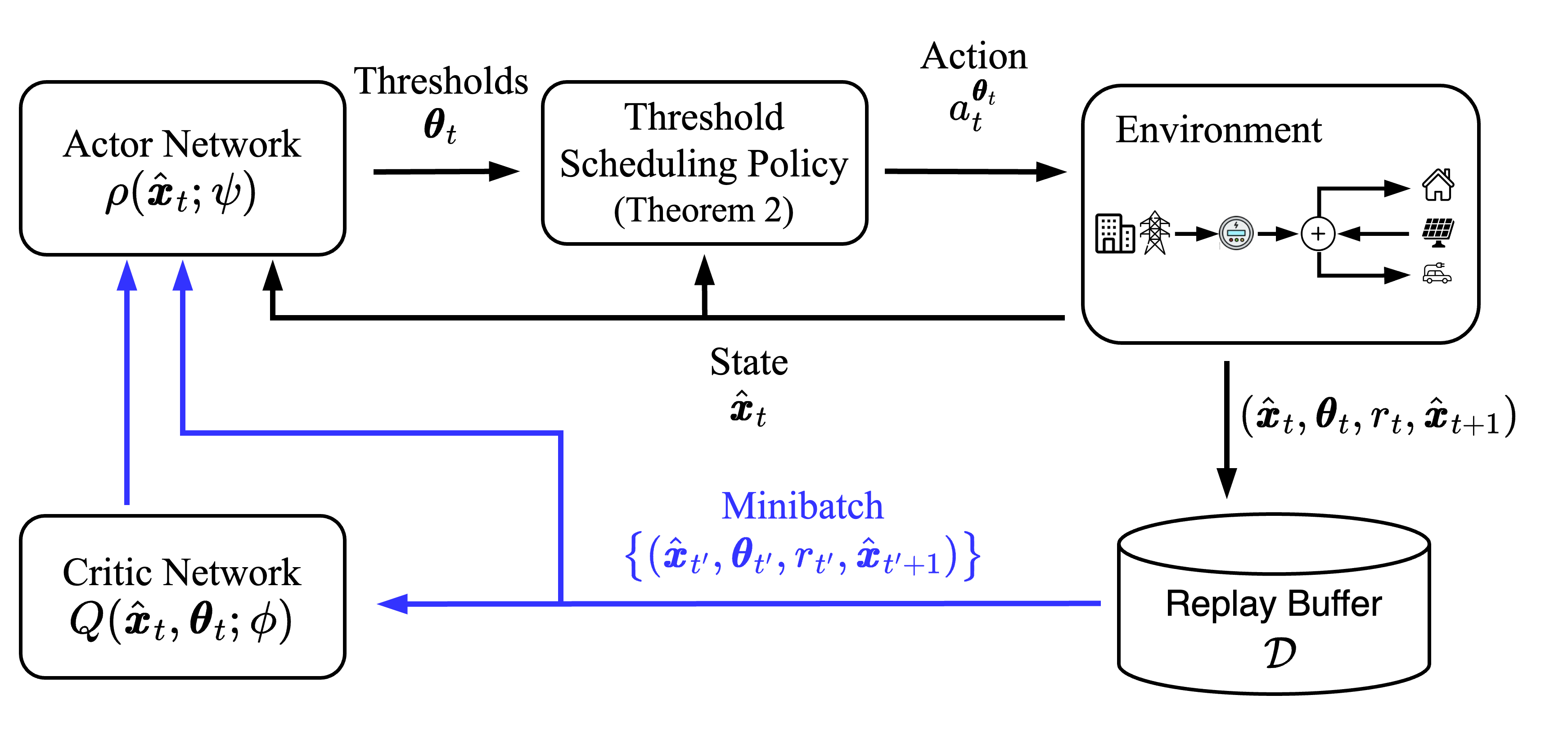}
		\caption{Threshold learning algorithm based on SAC method.}
		\label{fig:Learning diagram}
	\end{figure}

	The overview of PTRL is presented in Fig.~\ref{fig:Learning diagram}. PTRL uses two neural networks, the actor network $\rho(\hat {\bm x}_t ; \psi)$ and the critic network $Q(\hat {\bm x}_t, \bm \theta_t ; \phi)$, with the neural network parameters $\psi$ and $\phi$, respectively. The actor generates procrastination parameters $\bm \theta_t$ based on the current state $\hat {\bm x}_t$, which is then used to compute the schedule for deferrable and nondeferrable loads $\bm a_t^{\bm \theta_t}$. Then we observe the household's surplus $r_t$, and the next state $\hat {\bm x}_{t+1}$. The tuple of state, action, reward and next state is stored in the replay buffer, which is later used to update neural networks. The critic network takes state and procrastination thresholds as inputs and approximates the reward-to-go from that state-threshold pair (the $Q$-function).
	
	For the balanced exploration and exploitation, the SAC algorithm learns a stochastic policy that maximizes the cumulative reward with the entropy regularization \cite{haarnoja2018softarxiv}. Although the optimal policy in Theorem~\ref{thm:opt2} is deterministic, PTRL learns a stochastic policy that approximates the optimal policy. Here, we use the standard Gaussian reparameterization trick, in which the output of the actor network is the mean and covariance matrix of procrastination thresholds:
	\begin{equation} \label{eq:gaussianparam}
		\rho(\hat {\bm x}_t ; \psi) = (\bar{\bm \theta}_t (\psi), \Sigma_t(\psi)), \;
\bm \theta_t(\psi) = \bar{\bm \theta}_t (\psi) + \big(\Sigma_t(\psi)\big)^{\frac 1 2}\epsilon,
	\end{equation}
	where $\epsilon \sim \mathcal N(0,I)$, and $I$ is the identity matrix. 
		
	The two neural networks are optimized by stochastic optimization methods during the updates. The actor network is optimized by maximizing the critic function with the entropy of the policy:
	\begin{equation}
	 	\begin{aligned}
			L_\rho(\psi) = \mathbb E_{\hat {\bm x}_t\sim \mathcal D,\,\epsilon \sim \mathcal N(0,I)} &
			\bigg[Q(\hat {\bm x}_t, \bm \theta_t (\psi); \phi) \\
			 - \alpha \log \varphi & \left(\bm \theta_t(\psi) ;\bar{\bm \theta}_t(\psi), \Sigma_t(\psi)  \right)  \bigg],\label{eq:ActorLoss}
		\end{aligned}
	\end{equation}
	\begin{equation}
		\psi \leftarrow \psi  + \lambda_\psi L_\rho(\psi), \label{eq:ActorUpdate}
	\end{equation}
	where $\varphi\left(\bm \theta_t; \bar{\bm \theta}_t, \Sigma_t \right)$ is the PDF of Gaussian distribution with mean $\bar{\bm \theta}_t$ and covariance matrix $\Sigma_t$. The temperature coefficient	$\alpha$ is a parameter that controls the randomness of the optimal policy by balancing the trade-off between rewards and entropy.
	
	The critic network is optimized by minimizing the mean squared bellman error $L_Q(\phi)$:
	\begin{align}
		L_Q(\phi) &= \mathbb E_{(\hat {\bm x}_t,\bm \theta_t,r_t,\hat {\bm x}_{t+1}) \sim \mathcal D}
		\left[ \big( Q(\hat {\bm x}_t,\bm \theta_t;\phi ) - \xi(r_t, \hat {\bm x}_{t+1}) \big) ^2\right],\label{eq:MSBE} \\
		\phi &\leftarrow \phi - \lambda_\phi \nabla_\phi L_Q(\phi),\label{eq:criticupdate}
	\end{align}
	where $ \xi(r_t, \hat {\bm x}_{t+1})$ is the temporal difference target defined as:
		\begin{equation}\label{eq:BellmanTarget}
			\begin{aligned}
				\xi(r_t,\hat {\bm x}_{t+1})= r _t+   Q(\hat {\bm x}_{t+1}, &\bm \theta_{t+1}; \bar \phi ) \\ 
		- \alpha &\log \varphi  \left(\bm \theta_{t+1} ; \bar{\bm \theta}_{t+1}, \Sigma_{t+1}  \right). 
		\end{aligned}
		\end{equation}
		As in (\ref{eq:gaussianparam}), $\bm{\theta}_{t+1}, \bar{\bm \theta}_{t+1}$, and $\Sigma_{t+1}$ are obtained by evaluating the actor network $\rho$ at $\hat{\bm x}_{t+1}$. 
	
	The target critic network $Q\big(\hat {\bm{x}}_{t+1}, \bm \theta_{t+1}; \bar \phi \big)$ is another neural network which has the exact same network structure with the critic network and it is used for stabilizing learning. The parameter for the target critic network $\bar \phi$ is updated as:
	\begin{equation}
		\bar \phi \leftarrow \tau \bar \phi + (1 - \tau )\phi,
	\end{equation}
	for $\tau \in [0, 1]$.
	
	The expectations in (\ref{eq:ActorLoss}) and (\ref{eq:MSBE}) are approximated using the $N$ minibatch samples $\{(\hat {\bm x}_t, \bm \theta_t, r_t, \hat {\bm x}_{t+1}) \}$ sampled from the replay buffer $\mathcal D$. The detailed steps of the PTRL are described in Algorithm \ref{alg:ThresholdLearning}. 

    \begin{algorithm}[tbh]
    \caption{PTRL algorithm}\label{alg:ThresholdLearning}
    \begin{algorithmic}
    \State Initialize parameters $\phi, \psi$
    \State Initialize $\bar \phi$: $\bar \phi \leftarrow \phi$
    \Repeat
        \State Sample initial state $\hat {\bm x}_0$ and $T$
        \For{$t = 0, 1, \ldots, T$}
        	\State Compute $(\bar{\bm \theta}_t, \Sigma_t) = \rho (\hat {\bm x}_t ; \psi)$.
            \State Sample $\bm \theta_t = (\theta_t^-, \theta_t^0, \theta_t^+) $ using (\ref{eq:gaussianparam})
            \State Compute $\bm a_t^{\bm \theta_t}$ using Theorem 2.
            \State Observe next state $\hat {\bm x}_{t+1}$ and reward $r_t$
            \State Store $(\hat {\bm x}_t, \bm \theta_t, r_t, \hat {\bm x}_{t+1})$ in the replay buffer $\mathcal D$
            \If{it's time to update}
                    \State Randomly sample a minibatch from $\mathcal D$
                    \vspace{-0.4em}
                    \begin{equation*}
                        B = \{(\hat {\bm x}_{t'}, \bm \theta_{t'}, r_{t'}, \hat {\bm x}_{t'+1})\}\sim \mathcal D
                    \end{equation*}
                    \vspace{-1.6em}
                    \State Compute $\xi(r_{t'}, \hat {\bm x}_{t'})$ using (\ref{eq:BellmanTarget})
                    \State Update $\phi$ using (\ref{eq:MSBE}) and (\ref{eq:criticupdate})
                    \State Update $\psi$ using (\ref{eq:ActorLoss}) and (\ref{eq:ActorUpdate})
                    \State $\bar \phi \leftarrow \tau \bar \phi + (1 - \tau) \phi$
            \EndIf
        \EndFor
    \Until{convergence}
    \end{algorithmic}
    \end{algorithm}
    
        When $g_t$ is time independent, the procrastination thresholds satisfy following linear time dependency, further reducing the number of parameters to be learned.
    
	\begin{proposition}[Linear time dependency of procrastination thresholds] \label{prop:thresholdsproperty}
	If $g_t$ is time independent and $\pibf_t = \pibf_{t+1}$, the procrastination thresholds for net consuming and net producing modes, $\theta_t^+$ and $\theta_t^-$, satisfy linear time dependency:
	     	\begin{align*}
	     		\theta_{t}^- = \theta_{t+1}^-, \quad \theta_{t}^+ = \theta_{t+1}^+ + \bar v.
	     	\end{align*}
	\end{proposition}
	
	The linear time dependency of the procrastination thresholds implies that we only need to compute the thresholds for the last interval of each pricing period. This makes the number of thresholds to be learned independent of the scheduling horizon. To exploit this linear time dependency, we treat $\theta_t^-$ and $\theta_t^+$ for the last interval of each pricing period as scalar learnable parameters and update them directly via (\ref{eq:ActorLoss})--(\ref{eq:ActorUpdate}); their values at all other intervals are computed using the Proposition~\ref{prop:thresholdsproperty}. Since Proposition~\ref{prop:thresholdsproperty} does not apply to $\theta_t^0$, it remains an output of the actor network.

\section{Numerical Results}\label{sec:numericlaresults}
    In this section, we present the simulation results for the prototype model in Sec.~\ref{sec:problemformulation}, joint scheduling of deferrable and nondeferrable jobs in the household, comparing the performance of the proposed method with several benchmark deep reinforcement learning algorithms and other control methods. We first describe the simulation setting in Section~\ref{sec:setting}, followed by the numerical results.

 \subsection{Simulation setting} \label{sec:setting}
 \subsubsection{Evaluation datasets}
The evaluations of the proposed scheduling policy and comparison benchmarks were based on two real-world datasets. We used data collected from the Adaptive Charging Network (ACN) dataset \cite{lee_acndata_2019}, which provides charging demand, demand arrival time, and the deadline for each charging session. The evaluation also used the Pecan Street dataset~\cite{pecanstreetdata}, which provides rooftop solar generation and household electricity consumption data for houses in New York State. Both the reinforcement learning algorithm's training and performance testing were based on these datasets.

\subsubsection{Demand}
We considered a single household with nondeferrable and deferrable demands. The two types of nondeferrable demand were HVAC and standard consumption. Their utility functions were assumed to be quadratic, with parameters estimated from historical retail price and household electricity usage data in the Pecan Street dataset using the inverse optimization method of \cite{keshavarz2011imputing}. For simplicity, we assumed time-independent utility functions.

The deferrable demand was in the form of a sequence of EV charging requests defined by three random attributes: arrival time of the charging request, kWh demanded, and completion deadline; these parameters were assumed to be known at the time of the request. The maximum capacity of the charger was 7.2 kW, typical for a level-2 charger.

\subsubsection{Supply} The household demands were met by two sources:  the local renewable generation and grid-purchased electricity from the distribution utility. Renewable generation has zero marginal cost, whereas purchased electricity was subject to the Time-of-Use (ToU) NEM tariff of the distribution utility. In the simulation, we used the Pacific Gas and Electric (PG\&E) ToU retail rate with on-peak hours defined as 4 PM to 9 PM. The NEM import-export rates $(\pi_t^+,\pi_t^-)$ were set at $(0.57\,\$/\,\text{kWh}, \, 0.23\,\$/\,\text{kWh})$  during peak hours and $(0.43\,\$/\,\text{kWh}, 0.17\,\$/\,\text{kWh})$ during off-peak hours.

\subsubsection{Scheduling framework}  The arrival of each charging request initiated a charging episode beginning at the time of request and ending at the charging deadline, during which the deferrable and nondeferrable demands are co-optimized.\footnote{The reward from scheduling nondeferrable outside the episode was not included in the performance comparison.} Each episode consisted of 15-minute control intervals. The performance was evaluated by simulating a large number of EV charging episodes to obtain the sample average of cumulative rewards.

\subsubsection{Training  Strategy}
The optimal scheduling derived in Section~\ref{sec:thresholdpolicy} assumes Markovian DG, whereas the field-collected rooftop solar data may not satisfy this assumption. 

A significant limitation of implementing RL is the limited training data. A standard approach to circumvent data limitations is to use the available training data to learn a parametric generative model from which unlimited training data could be produced. So long as the test data are field-collected and statistically independent of the training data, the test results are justified. This two-stage approach—training on synthetic data from fitted models and testing on independent field data—also serves to verify the validity of the fitted models: consistent performance on out-of-sample field data confirms that the generative models capture sufficient structure for effective policy learning.

To this end, we used the historical data to fit two models for renewable generation. One was a first-order Gaussian-Markov (autoregressive) model and the other a time-independent model with a truncated Gaussian distribution, both with parameters estimated from training data.
 \begin{table}[t]
	\centering
	\caption{Hyperparameter setting}
	\begin{tabular}{cc}
		\toprule
		Hyperparameters & Value  \\
		\midrule
		Actor learning rate $(\lambda_\psi)$ & $5 \times 10^{-4}$  \\
		Batch size & 256 \\
		Buffer size & $10^5$ \\
		Critic learning rate $(\lambda_\phi)$ & $3 \times 10^{-3}$ \\
		Hidden layer size & [128, 128] \\
		Target smoothing coefficient $\tau$ & $5 \times 10^{-4}$\\
		\bottomrule
	\end{tabular}
	\label{tab:hyperparameters}
\end{table}

\subsubsection{Benchmark comparisons}
We compared the performance of PTRL, with three benchmarks: SAC \cite{haarnoja2018softarxiv}, DDPG \cite{lillicrap2015continuous}, and standard MPC solutions. SAC was selected as it shares the same underlying algorithm as PTRL, but learns over the full $(K+1)$-dimensional action space, isolating the benefit of the threshold structure. DDPG was included for its established use in continuous action EV charging problems \cite{ye2021model}. MPC serves as a widely adopted non-learning baseline.

All learning algorithms used multilayer perceptrons (MLPs) with two hidden layers, and the ReLU and hyperbolic tangent (tanh) functions as activation functions for the hidden and output layers, respectively. Neural networks for all learning algorithms were initialized with He uniform initialization \cite{he2015delving}. All learning algorithms were trained on the same training dataset and evaluated on the same test dataset to ensure a fair comparison.

Hyperparameters were tuned via grid search over learning rates and hidden layer sizes for DDPG, as it is the most sensitive to hyperparameter selection. The same values were applied to SAC and PTRL. For both SAC and PTRL, the temperature coefficient $\alpha$ was tuned following the method introduced in \cite{haarnoja2018softarxiv}. A hyperparameter sensitivity analysis for learning rate and hidden layer size is provided in Appendix~\ref{sec:appendixC}. Used hyperparameters are listed in Table~\ref{tab:hyperparameters}.

The MPC controller used a rolling prediction window of 20 intervals (5 hours), chosen to capture at least one transition between on-peak and off-peak pricing periods while maintaining reasonable computational cost. The renewable prediction within the window was the unconditional mean for the independent model and the conditional expectation from the fitted AR(1) model for the Markovian model, both of which are minimum mean squared error estimates.

\subsubsection{Performance Evaluation}
The performance of learning was evaluated in two ways; one was based on the out-of-sample synthetic data generated under the same model used in training. Such tests demonstrate the performance under ideal conditions, where the data model in training matches perfectly with that in testing.

A more realistic evaluation can be achieved by using field-collected data to assess the performance of RL solutions trained under independent or Markovian models. Such evaluations include the effects of model mismatch between training and actual real-world operation and robustness of the learning algorithms. 

\subsection{Learning Performance}\label{sec:VB}
This experiment was based on synthetic DG data traces generated from the Markovian and independent models. The goal was to test the learning performance of PTRL and related benchmarks. 

\begin{figure}[t]
	\begin{subfigure}[b]{0.49\columnwidth}
		\includegraphics[width=\linewidth]{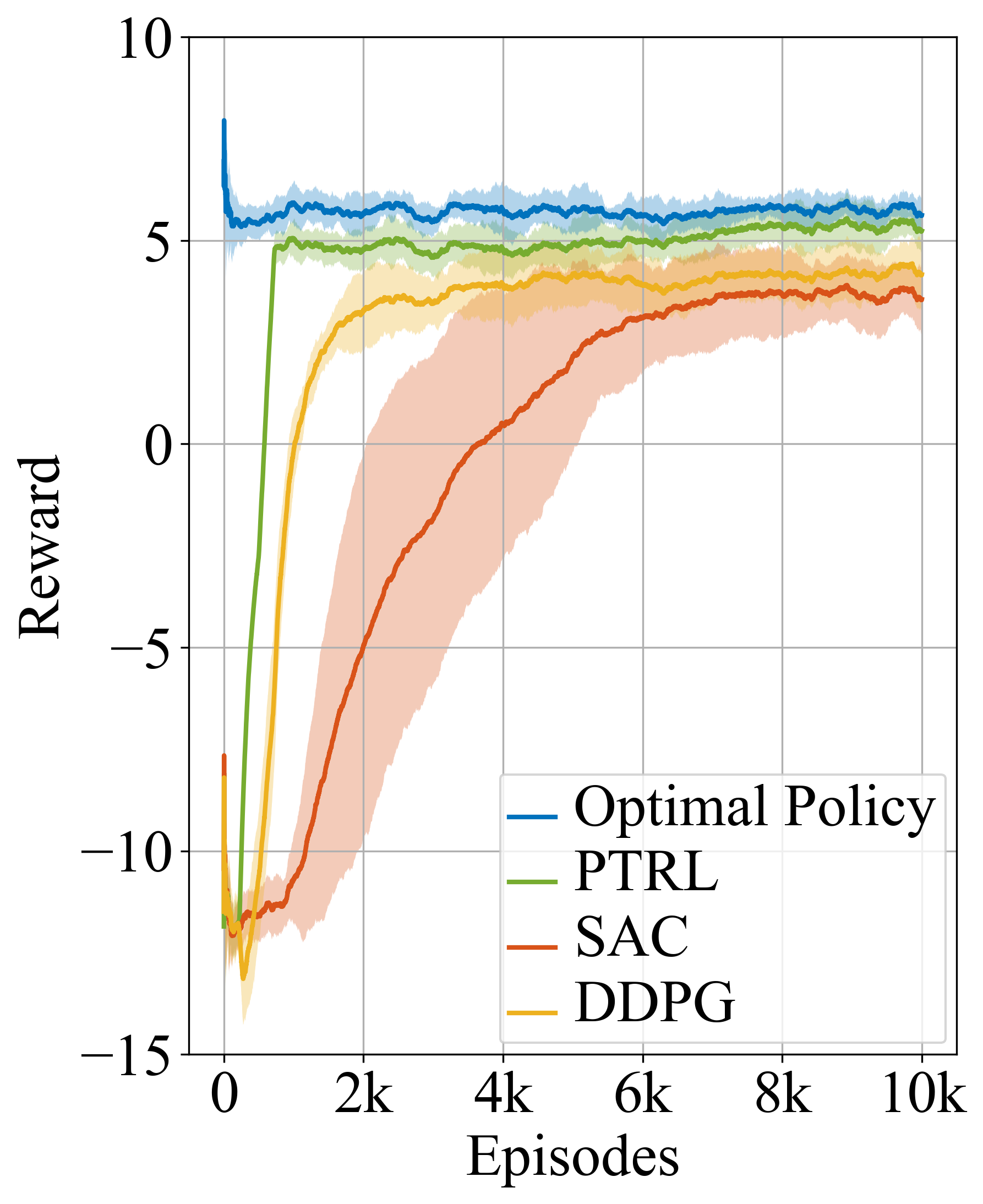}
		\caption{}
		\label{fig:LearningCurveIndependent}
	\end{subfigure}
	\hfill 
	\begin{subfigure}[b]{0.49\columnwidth}
		\includegraphics[width=\linewidth]{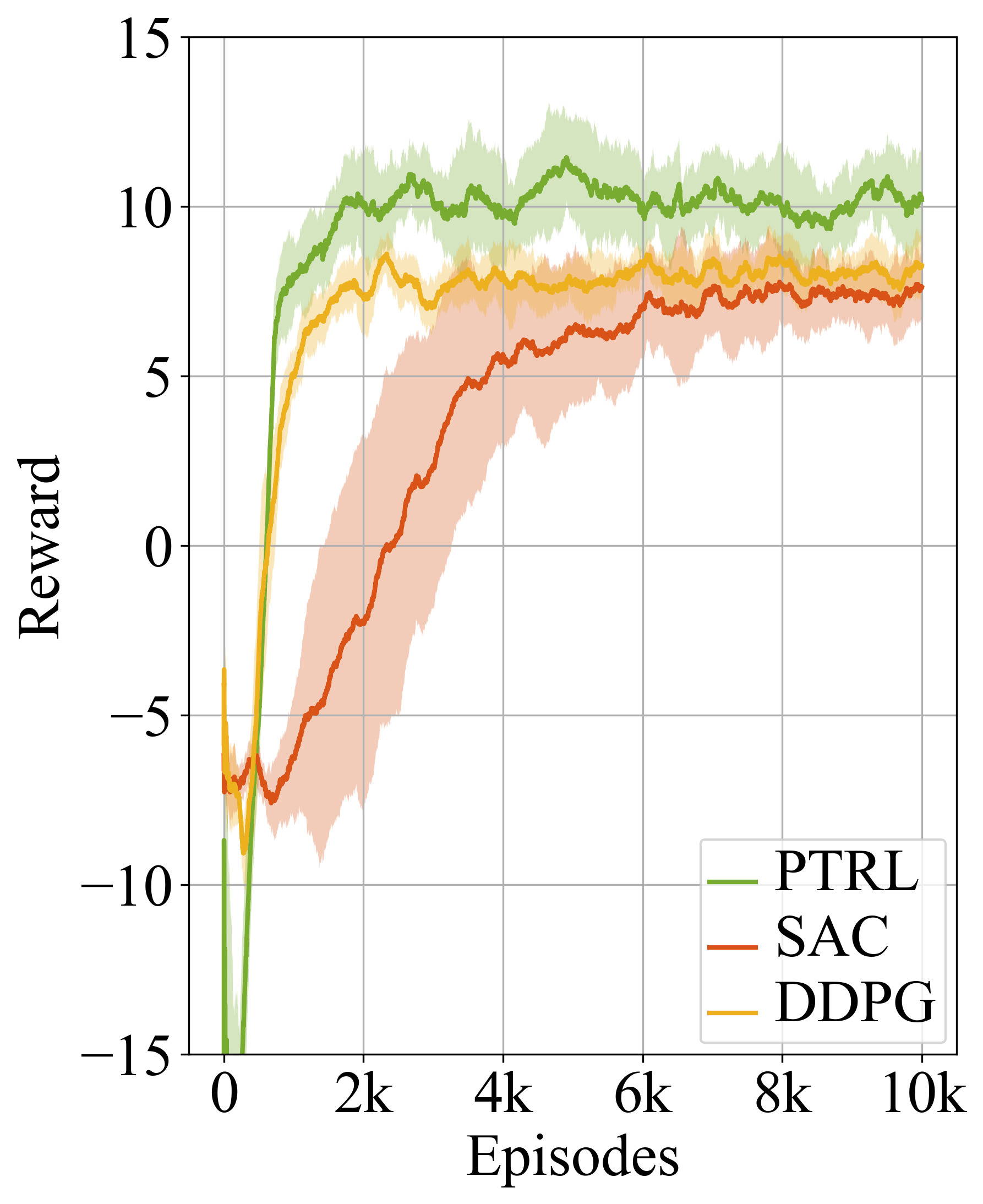}
		\caption{}
		\label{fig:LearningCurveMarkov}
	\end{subfigure}
	\caption{Learning curve of (a) independent Gaussian DG model (b) Markovian DG model}
	\label{fig:LearningCurve}
\end{figure}

Fig.~\ref{fig:LearningCurve} shows the learning curves of different learning algorithms. The left and right panels are the learning curves under independent and Markovian model, respectively. Each learning curve is the moving average of cumulated rewards over the last 200 episodes.

PTRL achieved the fastest convergence among the tested benchmarks under both data models. PTRL converged within approximately 800 episodes under the independent model and 2,000 episodes under the Markovian model. We attributed the faster convergence of the independent model to the use of the explicit recursion for computing procrastination parameters shown in Proposition~\ref{prop:thresholdsproperty}. Note further that, under the independent model, the optimal policy can be easily evaluated using Proposition~\ref{prop:thresholdsproperty}. See the dark blue curve in Fig. \ref{fig:LearningCurve} (a). The gap to optimality of PTRL under the independent model was within 6.80\%. The standard deviations (normalized by the mean) of PTRL at the end of training were 0.1036 under the independent model and 0.0763 under the Markovian model.

SAC performed similarly under both models, converging roughly at 6,000 episodes with normalized standard deviation at 0.2386 under the independent model and 0.125 under the Markovian model. The gap to optimality under the independent model was 36.76\%. The DDPG also performed similar under both models, converging roughly at 2,000 episodes with a normalized standard deviation at 0.2060 under the independent model and 0.1233 under the Markovian model. The gap to optimality under the independent model was 25.91\%. The performance of DDPG under both models was better than the SAC.

\begin{figure}[t]
	\centering
	\includegraphics[width =0.85 \columnwidth]{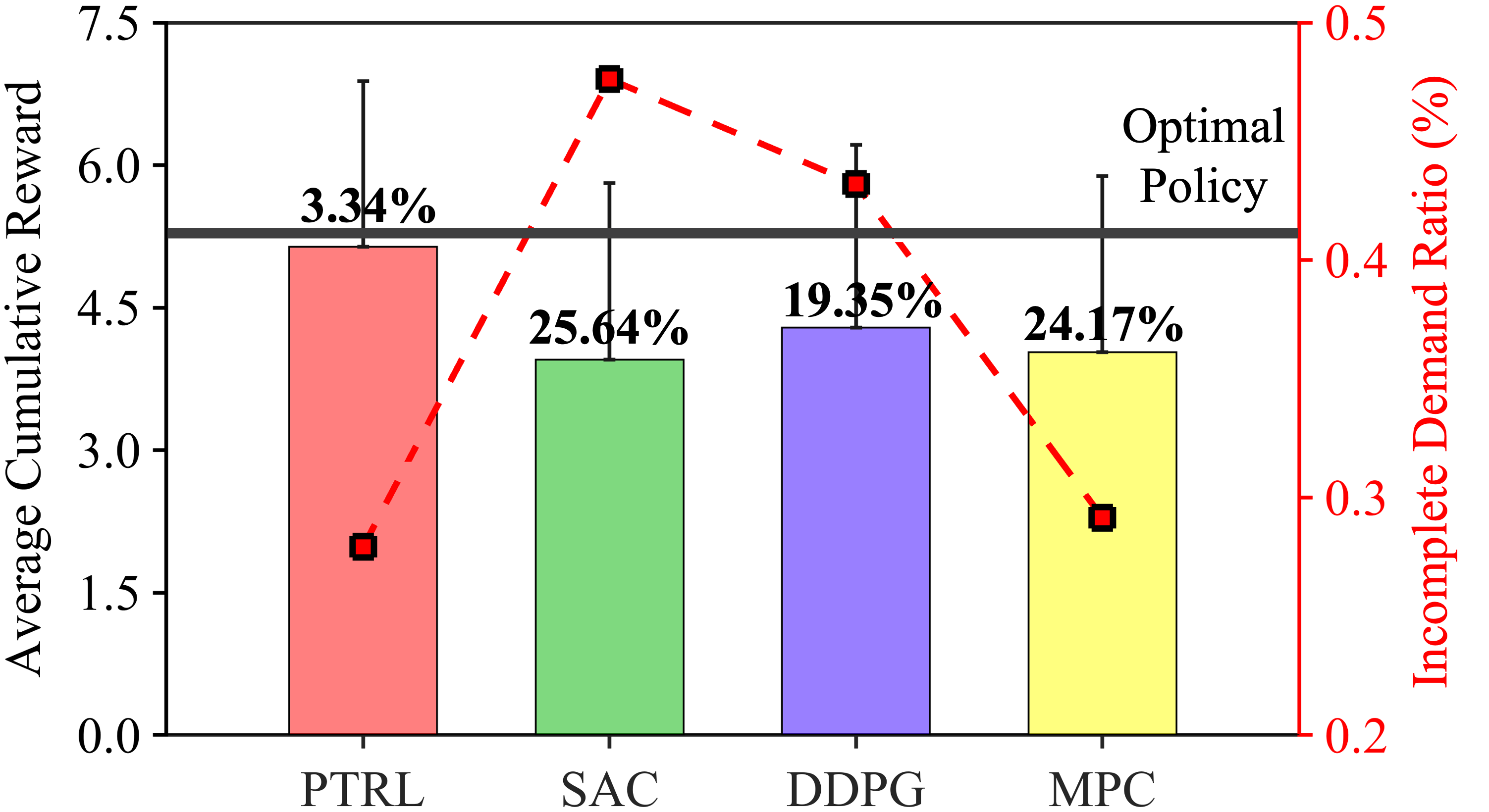}
	\caption{Average cumulative reward under synthetic test dataset of time independent DG model. Error bars indicate standard deviations, percentages show relative gap to optimal, and the red dashed line shows incomplete demand ratio.}
	\label{fig:MCRun_ind_syn}
\end{figure}

\begin{figure}[t]
	\centering
	\includegraphics[width =0.85 \columnwidth]{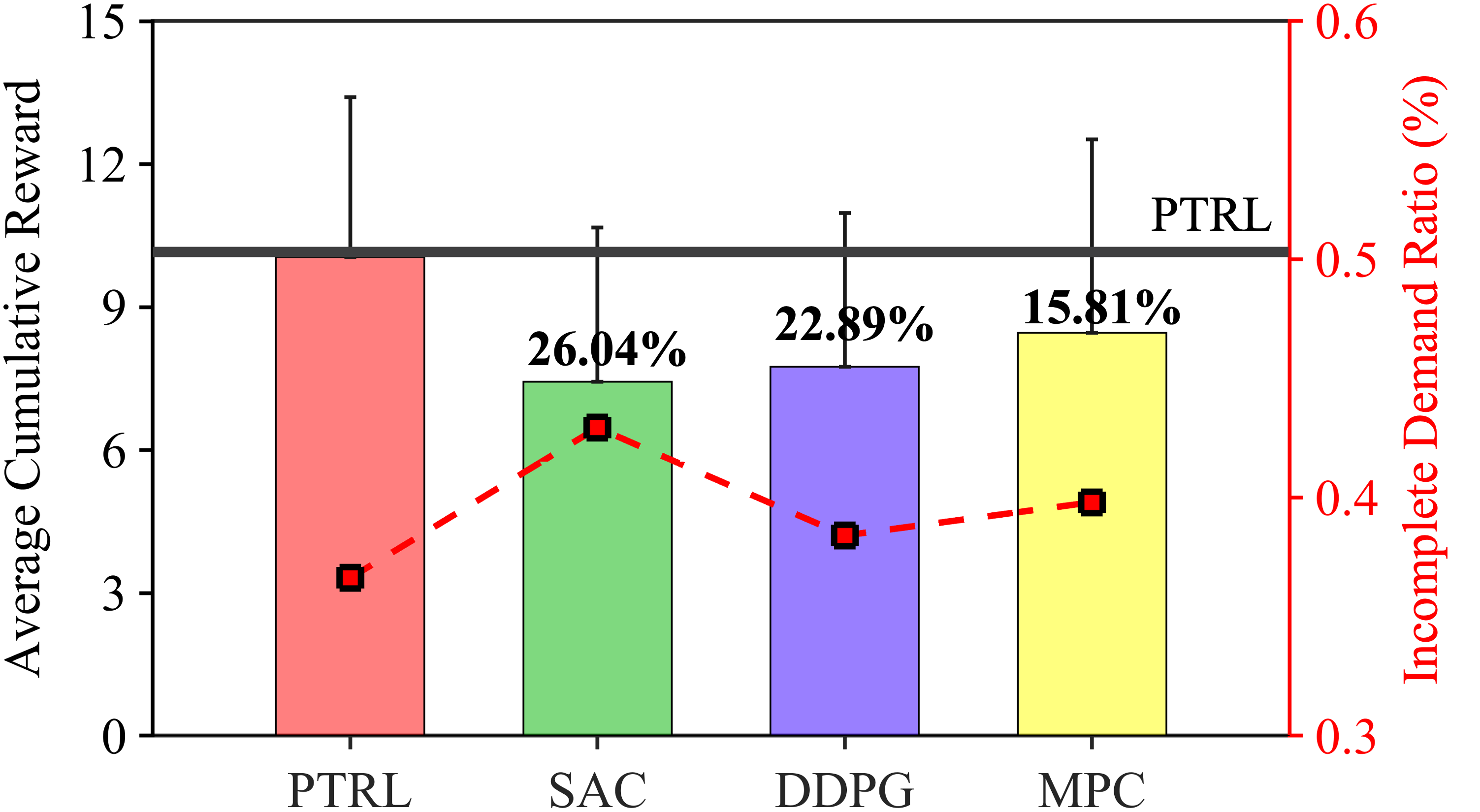}
	\caption{Average cumulative reward under synthetic test dataset of Markovian DG model. Error bars indicate standard deviations, percentages show relative gap to PTRL, and the red dashed line shows incomplete demand ratio.}
	\label{fig:MCRun_Mark_syn}
\end{figure}

In Figs.~\ref{fig:MCRun_ind_syn} and \ref{fig:MCRun_Mark_syn}, we plotted the Monte Carlo simulation results using the synthetic test dataset that follows the same distribution as the training dataset. PTRL had the smallest gap to the optimal policy, achieving 3.34\%, which aligns with the learning results. Under Markovian model, we could not compute optimal policy, hence, the relative gap to the PTRL algorithm is shown in Fig.~\ref{fig:MCRun_Mark_syn}. The MPC's performance improved under Markovian model, which was possible due to better prediction using AR(1) model. In addition to cumulative rewards, Figs.~\ref{fig:MCRun_ind_syn} and \ref{fig:MCRun_Mark_syn} report the average incomplete demand ratio (red dashed line) and the standard deviation of rewards (error bars). PTRL achieved the lowest incomplete demand ratio among all benchmarks. The standard deviations were comparable across all algorithms, suggesting that reward variance was primarily driven by the stochasticity of EV charging demand arrivals and renewable generation rather than by the scheduling algorithm.

\begin{figure}[t]
	\centering
	\includegraphics[width =0.7\columnwidth]{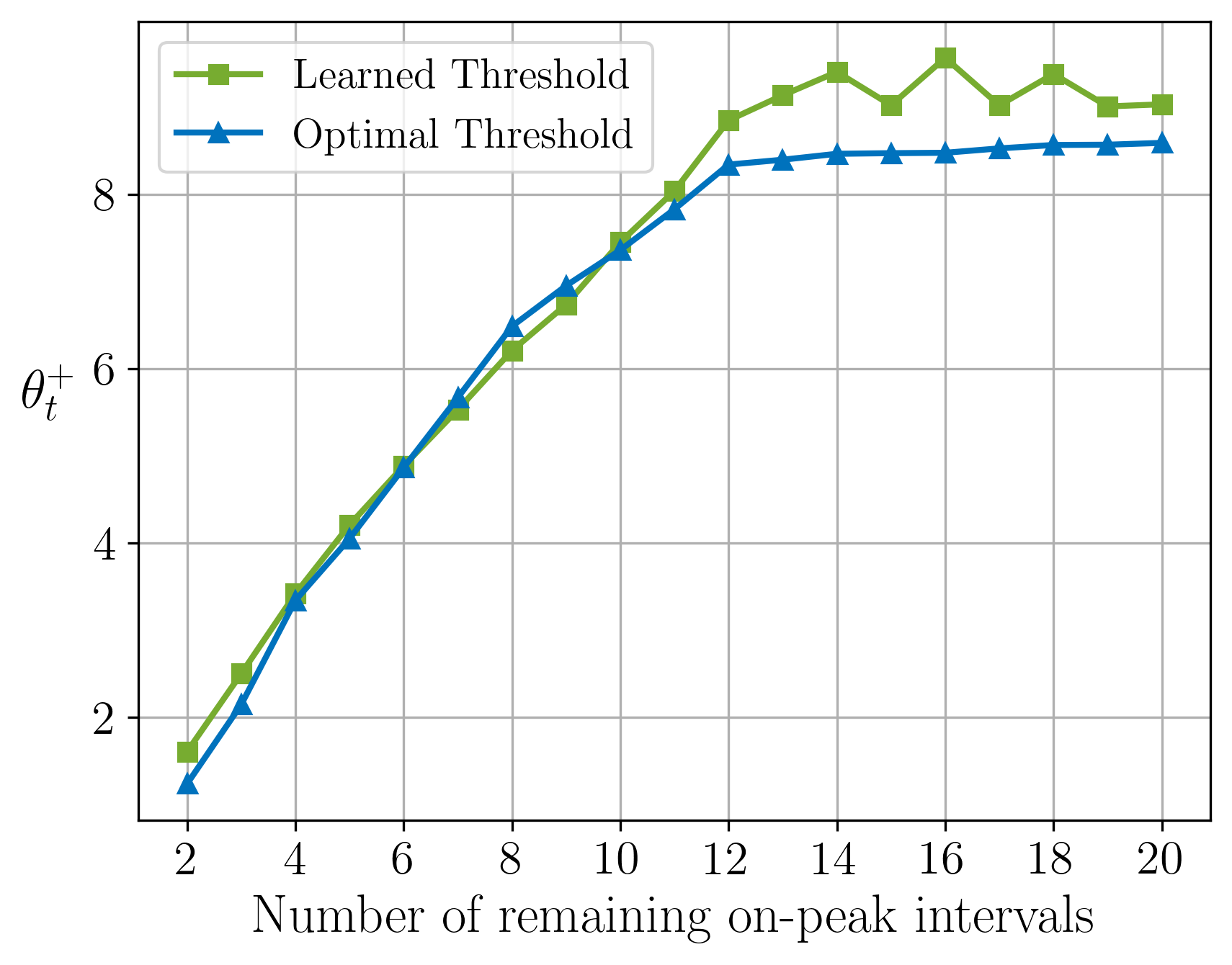}
	\caption{Learned vs. optimal procrastination thresholds $\theta_t^+$ under the independent renewable generation model.}
	\label{fig:learned_threshold}
\end{figure}

Fig.~\ref{fig:learned_threshold} compares the thresholds learned by PTRL with the optimal thresholds at the last interval of the off-peak period, under the independent renewable generation model. The upper procrastination threshold $\theta_t^+$ was plotted against the number of remaining on-peak intervals following the off-peak period. The learned thresholds closely trackd the theoretical values across most of the range. Excluding the first two data points, where the absolute values of the optimal thresholds are small and relative errors are amplified, the learned thresholds fell within 0.42--10.90\% of the optimal values.

The computation times for evaluating 2000 Monte Carlo runs were 779.8 s for MPC, 42.0 s for PTRL, 10.4 s for SAC, and 4.7 s for DDPG. MPC was approximately 19 times slower than PTRL, demonstrating the computational benefit of PTRL in the online implementation.

\subsection{Evaluation with Field Data}
\begin{figure}[t]
	\centering
	\includegraphics[width =0.7 \columnwidth]{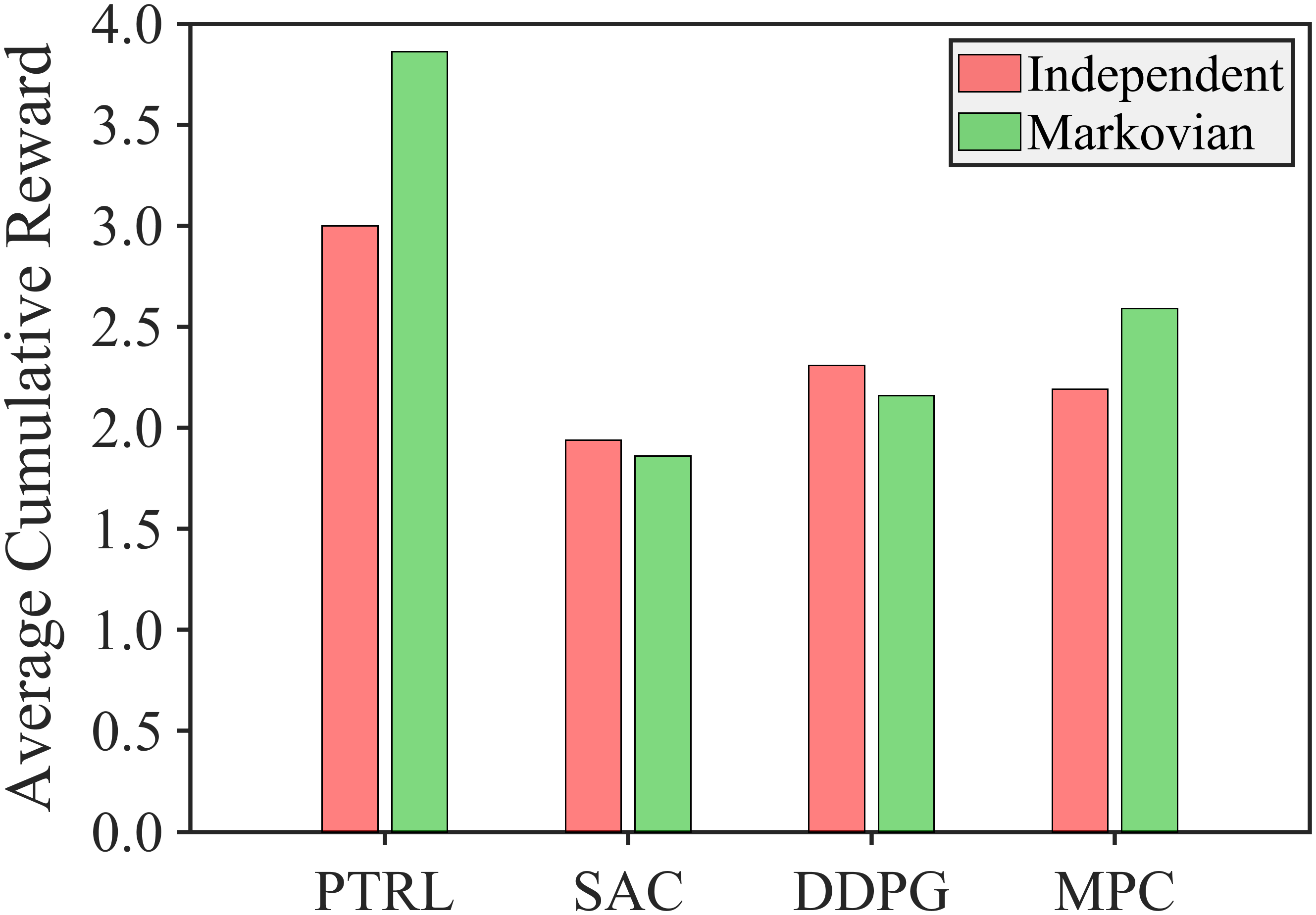}
	\caption{Average cumulative reward of 400 Monte Carlo runs using field data from Pecan Street dataset.}
	\label{fig:MCRun_Act}
\end{figure}

We plotted the Monte Carlo run results using the field-collected dataset from the Pecan Street dataset in Fig.~\ref{fig:MCRun_Act}. The proposed method outperformed all benchmark scheduling algorithms in both independent and Markovian models, indicating PTRL was more robust to the randomness in renewables than other benchmarks. PTRL outperformed SAC and DDPG by about 107.4\% and 79.0\%, respectively. These results verified the fitted generative models used in training. Despite the mismatch between the Markovian assumption and actual field data, PTRL consistently outperformed all benchmarks, confirming that the approach is robust under real-world conditions. One thing to note was that the average cumulative reward of the PTRL trained with the Markovian model was about 28.0\% higher than that of the independent model. This suggests that the actual DG is closer to the Markovian model. 

\subsection{Deferrable load scheduling simulation results}
To better understand the benefit of procrastination policy and the performance of the PTRL, we applied it to a simplified scenario in which only deferrable loads are scheduled. The simulation setup was identical to the general case, except that only the deferrable demand was served either by DG or by imported power from the grid. For training, we used a Markovian model for local renewable generation.

\begin{figure}[t]
	\centering
	\includegraphics[width =0.85\columnwidth]{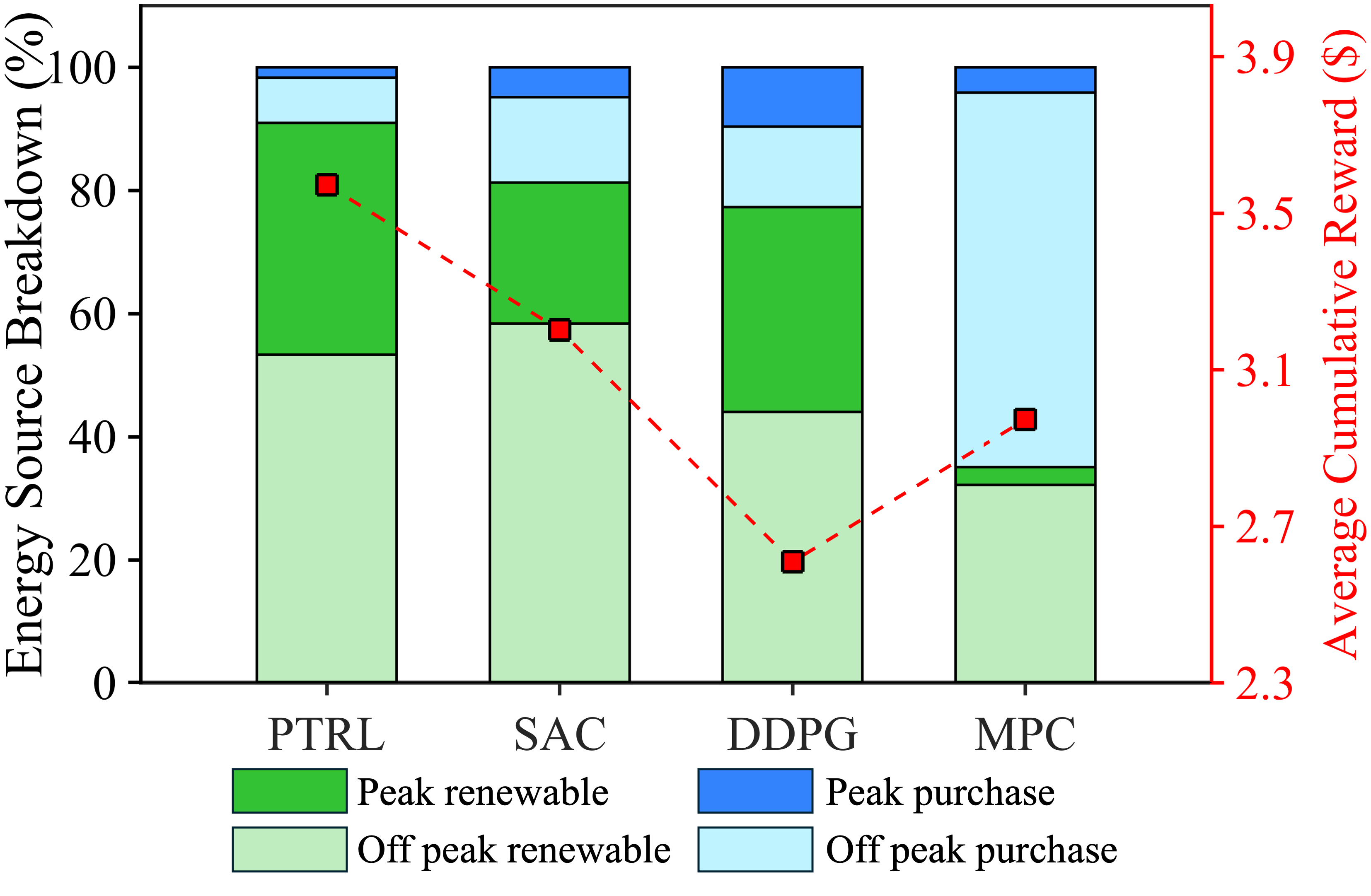}
	\caption{Energy source breakdown of the completed deferrable loads and average cumulative reward evaluated with 400 field dataset from Pecan Street.}
	\label{fig:energymix}
\end{figure}

We plotted average cumulative reward and energy source breakdown of completed deferrable loads evaluated using the actual field data after training in Fig.~\ref{fig:energymix}. The energy source breakdown shows the percentage of energy used to complete the deferrable demand by source (local renewables vs. grid purchases) and pricing period (peak vs. off-peak). Compared to benchmark algorithms, PTRL achieved the highest cumulative reward, primarily due to its high utilization of local renewable generation (the sum of the dark and light green bars). While the MPC algorithm yielded a higher reward than DDPG despite using less local renewable generation, this was mainly because it served most of its load using grid electricity during off-peak periods, the light blue bar, thereby reducing purchases in peak periods, the dark blue bar.

Regarding the use of renewables, it is preferable to consume renewable generation during off-peak hours, when the export price is lower $(\pi_{\text{on}}^- > \pi_{\text{off}}^- )$, thereby reserving higher-value peak-hour renewables for export. In this figure, this is represented by a taller light green bar relative to the dark green bar. The energy source breakdown of PTRL indicates that it learned to preferentially consume renewables during off-peak hours. Similarly, SAC also exhibited this strategy, but it resulted in lower rewards than PTRL. This was primarily because SAC, lacking the threshold structure, did not cap its service rate at the available DG level and consequently incurred grid purchases during the off-peak period.

\section{Conclusion}\label{sec:conclusion_v10}
This paper studies the joint scheduling of deferrable and nondeferrable jobs with colocated stochastic supply under nonlinear pricing of electricity, which arises broadly in the energy system applications including EV charging stations and data centers. We establish a low-dimensional parameterization of an optimal scheduling policy based on the procrastination principle, and develop a structured reinforcement learning algorithm (PTRL) to learn the optimal procrastination parameters. Together, these results transform an infinite-dimensional continuous-state MDP into a tractable learning problem.

One natural direction is relaxing the known utility assumption by incorporating online inverse optimization or imitation learning into the reinforcement learning algorithm, so that the utility of nondeferrable demands is learned jointly with scheduling decisions. Another interesting direction is extending the framework to multiple simultaneous deferrable tasks with different arrival times and deadlines, which calls for a low-dimensional parameterization that remains tractable as the number of tasks grows.
\appendices
\section{Proof of Theorems}\label{sec:appendixproof}
We use the following notations:
\begin{itemize}
	\item $\bar V_t(y_t) := \mathbb E_{g}[V_t(y_t, g)]$.
		\smallskip
	\item $\bar V_{t}(y_t, g_t) := \mathbb E_{g'}[V_t(y_t, g')\, | \,g_t]$.
		\smallskip
	\item $\partial^+_y\bar V_t(y), \, \partial ^- _y \bar V_t(y)$: Right and left derivative of $\bar V_t(y_t)$ at $y_t = y$, respectively.
	\smallskip
	\item $\partial^+_y \bar V_{t}(y; g), \, \partial^-_y \bar V_{t}(y;g)$: Right and left derivative of $\bar V_{t}(y_t, g)$ at $y_t = y$, respectively.
\end{itemize}

We use following two lemmas to prove both theorems. 

\begin{lemma} \label{lemma:concavity}
	The optimal value function $V_t(y_t, g_t)$ is a concave function of $y_t$ for all $t$.
\end{lemma}

\begin{proof}

	We prove the lemma by induction. For $t = T$,  the Bellman equation (\ref{eq:bellmanequation}) for $\bm x_T = (y_T, g_T)$ is 
	\begin{equation} \label{eq:apd1}
	\max_{v, \bm d} \quad  U(\bm d) - P_{\pi_T}(\mathbf{1}^T \bm d+ v - g_T) - q(y_T - v).
	\end{equation}
	The derivative of the objective function with respect to $v$ is $-\pi_T^+ + q'(y_T- v) > 0$ or $-\pi_T^- + q'(y_T - v) > 0 $, which are both positive by Assumption A3. Hence, the optimal deferrable demand service rate is its upper limit, $v_T^* = \min\{y_T, \bar v\}$. Since $P_{\pi_T}$ and $q$ are both convex functions of $v$, $V_T$ is concave in $y_T$. 

	Assume $V_{t+1}$ is concave in $y_{t+1}$. Since expectation preserves concavity, $\bar V_{t+1}(y_{t+1}, g_t)$ is also concave in $y_{t+1}$. The Bellman equation (\ref{eq:bellmanequation}) at $t$ therefore maximizes a concave objective over a compact, convex feasible set, and an optimal solution exists for every $(y_t, g_t)$.
	
	Fix $g_t$ and consider two states $\bm x_{t, 1} = (y_{t, 1}, g_t)$ and $\bm x_{t, 2} = (y_{t,2}, g_t)$ with optimal actions $(\bm d_{t,1}^*, v_{t,1}^*)$ and $(\bm d_{t,2}^*, v_{t,2}^*)$, respectively. For $\lambda \in [0, 1]$, define the convex combinations
	\begin{align*}
	\bm x_\lambda &= \lambda \bm x_{t,1} + (1 - \lambda) \bm x_{t,2}, \; 
	v_\lambda = \lambda v_{t,1}^* + (1 - \lambda)  v_{t,2}^*, \; \\
	\bm d_\lambda &= \lambda \bm d_{t,1}^* + (1 - \lambda)  \bm d_{t,2}^*.
	\end{align*}
By convexity of the feasible set, $(\bm d_\lambda, v_\lambda)$ is feasible at state $\bm x_\lambda$. Let $(\bm d_\lambda^*, v_\lambda^*)$ denote the optimal action at $\bm x_\lambda$. Then
	\begin{align*}
		\lambda V_t&(\bm x_{t,1}) + (1 - \lambda) V_t(\bm x_{t,2}) \\
		&= \lambda \left\{r_t \big( \bm x_{t,1}, (\bm d_{t,1}^*, v_{t,1}^*)\big) + \bar V_{t+1}(y_{t,1} - v_{t,1}^*, g_t)\right\} \\
		&+(1 - \lambda)\left\{r_t \big(\bm x_{t,2}, (\bm d_{t,2}^*, v_{t,2}^*)\big) + \bar V_{t+1}(y_{t,2} - v_{t,2}^*, g_t)\right\}  \\
		&\le r_t\big(\bm x_\lambda ,(\bm d_\lambda, v_\lambda) \big) + \bar V_{t+1}(y_\lambda - v_\lambda, g_t ) \\
		&\le r_t\big(\bm x_\lambda, (\bm d_\lambda^*, v_\lambda ^*) \big) + \bar V_{t+1}(y_\lambda - v_\lambda ^*, g_t) = V_t(\bm x_\lambda),
	\end{align*}
	where the first inequality follows from the concavity of $r_t$ and $\bar V_{t+1}$, and the second from the optimality of $(\bm d_\lambda^*, v_\lambda^*)$. Hence $V_t$ is concave in $y_t$ for all $t$.
\end{proof}

The following lemma presents the optimal policy for the remaining demand level in which the penalty is inevitable.

\begin{lemma} \label{lemma:optimal_penaltyregime}
	For $y_t \ge (T-t + 1)\bar v$, the optimal charging action is $v_t ^* = \bar v.$
\end{lemma}

\begin{proof}
	We prove by contradiction. Suppose $v_t^* < \bar v$ for some state with $y_t \ge (T-t+1)\bar v$. For the fixed realization $(g_t, \ldots, g_T)$, let $y_{T+1}^* > 0$ denote the resulting terminal remaining demand under the optimal policy. Charging $\epsilon > 0$ more at $t$ increases the stage cost by at most $\pi_t^+\epsilon$ but reduces the terminal penalty by $q(y_{T+1}^*) - q(y_{T+1}^* - \epsilon) > \pi_t^+\epsilon$ by Assumption A3. Since this holds for every sample path of DG and the DG distribution is independent of actions, $v_t^*$ cannot be optimal. 
\end{proof}

\subsection{Proof of Theorem 2}
	By Lemma~\ref{lemma:optimal_penaltyregime}, it suffices to consider $y_t \le (T-t+1) \bar v$. By Lemma~\ref{lemma:concavity}, the optimal value function $V_t$ is concave in $y_t$ and $\bar V_{t+1}(y_{t+1}, g_{t+1})$ is also concave in $y_{t+1}$ as expectation preserves concavity. Hence, the Bellman equation (\ref{eq:bellmanequation}) is a convex optimization, and both right and left derivatives of $\bar V_{t+1}$ with respect to $y_{t+1}$ exist.
	
	The NEM payment function $P_{\pi_t}(z_t)$ introduces a piecewise structure depending on the sign of the net consumption $v_t + \mathbf 1 ^T \bm d_t - g_t$. We solve the Bellman equation (\ref{eq:bellmanequation}) by partitioning into three cases for a fixed value of $g_t$: 1) net-consuming: $g_t < v_t + \mathbf 1^T \bm d_t$, 2) net-producing: $g_t > v_t + \mathbf 1 ^T \bm d_t$, and 3) net-zero: $g_t = v_t + \mathbf 1 ^T \bm d_t$. 
	
	\subsubsection{$g_t < v_t + \mathbf 1 ^T  \pmb d_t$ (Net-consuming)} The Bellman equation (\ref{eq:bellmanequation}) becomes
 \begin{align} \label{eq:apd2}	
		V_t\big((y_t, g_t)\big) = &\max_{\bm a \in \mathcal A, v + \mathbf 1^T \bm d > g_t} \; \Big(
		U_t(\bm d) - \pi_t^+(v + \mathbf 1^T \bm d - g_t) \nonumber \\
		&+ \mathbb E[V_{t+1}\big((y_t - v, g_{t+1})\big)\mid g_t]\Big).
	\end{align}
	For the nondeferrable demand, by the first order condition with respect to $\bm d$ applied to (\ref{eq:apd2}), the optimal schedule is
	\begin{equation*}
		d_{it}^* = d_{it}^+ := \min \left\{ \bar d_i, \partial U_{it}^{-1}(\pi_t^+)\right \} \quad \forall i = 1, \ldots, K.
	\end{equation*}
	
	For the deferrable demand, the first order conditions with respect to $v$ on (\ref{eq:apd2}) are:
	\begin{align}
		-\pi_t^+ - \partial_y^- \bar V_{t+1}(y_t - v; g_{t}) \le 0, \; (\text{if } v \ge 0), \label{eq:foc1} \\
		-\pi_t^+ - \partial_y^+ \bar V_{t+1}(y_t - v; g_{t}) \ge 0, \; (\text{if } v > 0). \label{eq:foc2}
	\end{align}
	For $\theta_t^+(g_t)$ that satisfies
		\begin{equation}\label{eq:thetaplus}
		\partial_y^+ \bar V_{t+1}(\theta_t^+ (g_t); g_t) \le -\pi_t^+ \le 	\partial_y^- \bar V_{t+1}(\theta_t^+(g_t) ; g_t),
	\end{equation}
	if $y_t < \theta_t^+(g_t)$, by the concavity of $\bar V_{t+1}$, its left derivative is nonincreasing: 
	\[
	-\pi_t^+ \le \partial_y^- \bar V_{t+1}(\theta_t^+(g_t); g_t) \le \partial_y^- \bar V_{t+1}(y_t; g_t).
	\]
	Hence, $v_t^* = 0$ satisfies the first order condition (\ref{eq:foc1}).
	
	If $y_t > \theta_t^+(g_t)$, $v_t^* = \min\{\bar v, y_t - \theta_t^+(g_t)\}$ satisfies the first order condition (\ref{eq:foc2}) by the monotonicity of $\partial_y^+\bar V_{t+1}$:
	\[
	\partial_y^+ \bar V_{t+1}(y_t - v_t^*; g_t) \le \partial_y^+ \bar V_{t+1}(\theta_t^+(g_t) ; g_t) \le -\pi_t^+. 
	\]
	In summary, for $\theta_t^+(g_t)$ that satisfies (\ref{eq:thetaplus}), $v_t^* = \min\{\bar v, [y_t - \theta_t^+(g_t)]^+\}$ is optimal.	
There are two edge cases where there does not exists $\theta_t^+(g_t)$ that satisfies (\ref{eq:thetaplus}. 

		\begin{enumerate}
		\item \underline{Case 1}: $-\pi_t^+ - \partial_y^+ \bar V_{t+1}(y_{t+1}; g_t) \ge 0$ for all $y_{t+1} \le (T-t)\bar v$. For all $y_t \le (T-t+1) \bar v$, $v_t^* = \min\{\bar v, y_t\}$ satisfies the first order condition (\ref{eq:foc2}) and is therefore optimal, corresponding to $\theta_t^+(g_t) = 0$.
		\item \underline{Case 2}: $ -\pi_t^+ - \partial_y^- \bar V_{t+1}(y_{t+1}; g_t) \le 0$ for all $y_{t+1} \le (T-t)\bar v$. For $y_t \le (T-t)\bar v$, $v_t^* = 0$ satisfies the first order condition (\ref{eq:foc1}) and is therefore optimal.\\
		For $y_{t+1} \ge (T-t)\bar v$, as the penalty is unavoidable for all $t' > t+1$, and by Lemma~\ref{lemma:optimal_penaltyregime}, $v_{t'}^* = \bar v$ for all $t' > t+1$. Then, the terminal remaining demand is strictly positive $y_{T+1} = y_{t+1} - (T-t)\bar v$, which implies that $\partial_y^+ \bar V_{t+1}(y_{t+1}; g_t) = -q'(y_{T+1})$ because the marginal reduction in $y_{t+1}$ directly reduces the penalty. By assumption A3,
		\begin{equation}\label{eq:apd4}
			 -\pi_t^+ - \partial_y^+ \bar V_{t+1}(y_{t+1}; g_t)  > 0 ,  \; \forall y_{t+1} \ge (T-t)\bar v.
		\end{equation}
		For $y_t \ge (T-t)\bar v$, $v_t^* = y_t - (T-t)\bar v$ satisfies the first order condition (\ref{eq:foc2}). Together, $v_t^* = \min\{\bar v, [y_t - (T-t)\bar v]^+\}$, with $\theta_t^+(g_t) = (T-t)\bar v$.
	\end{enumerate}
	\noindent{To summarize, for $g_t < \sum_{i=1}^K d_{it}^+ + \min \left\{ \bar v, [y_t - \theta_t^+(g_t)]^+ \right\}$, }
	\begin{equation*}
		(\bm d_t^*, v_t^*) = \left(\bm d_t^+, \min \left\{ \bar v, [y_t - \theta_t^+(g_t) ]^+ \right \} \right).
	\end{equation*}

	\subsubsection{$g_t > v_t + \mathbf 1 ^T \pmb d_t$ (Net-producing)} The Bellman equation (\ref{eq:bellmanequation}) becomes
	\begin{align} \label{eq:apd3}	
			V_t\big((y_t, g_t)\big) = &\max_{\bm a \in \mathcal A, v + \mathbf 1^T \bm d < g_t} \; \Big(
			U_t(\bm d) - \pi_t^-(v + \mathbf 1^T \bm d - g_t) \nonumber \\
			&+ \mathbb E[V_{t+1}\big((y_t - v, g_{t+1})\big)\mid g_t]\Big).
	\end{align}
	
For the nondeferrable demand, by the first order condition with respect to $\bm d$ applied to (\ref{eq:apd3}), the optimal schedule is
	\begin{equation*}
		d_{it}^* = d_{it}^- := \min \{ \bar d_i, \partial U_{it}^{-1}(\pi_t^-)\} \quad \forall i = 1, \ldots, K.
	\end{equation*}
	
	For the deferrable demand, the proof is parallel to the net-consuming case with $\pi_t^-$ replacing $\pi_t^+$. The first order conditions with respect to $v$ on (\ref{eq:apd3}) are:
		\begin{align}
		-\pi_t^- - \partial_y^- \bar V_{t+1}(y_t - v; g_{t}) \le 0, \; (\text{if } v \ge 0), \label{eq:foc3} \\
		-\pi_t^- - \partial_y^+ \bar V_{t+1}(y_t - v; g_{t}) \ge 0, \; (\text{if } v > 0). \label{eq:foc4}
	\end{align}
	For $\theta_t^-(g_t)$ that satisfies
	\begin{equation}\label{eq:thetaminus}
		\partial_y^+ \bar V_{t+1}(\theta_t^- (g_t); g_t) \le -\pi_t^- \le 	\partial_y^- \bar V_{t+1}(\theta_t^-(g_t) ; g_t),
	\end{equation}
	by the monotonicity of $\partial_y^{\pm}\bar V_{t+1}$ and the same edge case arguments as in the net-consuming case, 
	\[
	v_t^* = \min\{\bar v, [y_t - \theta_t^-(g_t)]^+\},
	\]
	satisfies (\ref{eq:foc3}) and (\ref{eq:foc4}) and is therefore optimal.

	{\noindent To summarize, for $g_t > \sum_{i=1}^Kd_{it}^- +\min \left\{ \bar v,[ y_t - \theta_t^- (g_t)]^+ \right\}$}, 
	\begin{equation*}
		(\bm d_t^*, v_t^*) = \left(\bm d_t^-, \min \left\{ \bar v,[ y_t - \theta_t^- (g_t)]^+ \right\}\right).
	\end{equation*}
	\subsubsection{$g_t = v_t + \mathbf 1^T \pmb d_t$ (Net-zero)} 
	We solve (\ref{eq:bellmanequation}) with the constraint $g_t = v_t + \mathbf 1^T \bm d_t$. The problem becomes
	\begin{equation}\label{eq:NZ_problem}
		\max_{(\bm d, v) \in \mathcal A, v + \mathbf 1^T \bm d = g_t} \; U_t(\bm d) + \bar V_{t+1}(y_t - v; g_t). 
	\end{equation}
	Since the optimization problem above satisfies the Slater's condition, the KKT conditions are necessary and sufficient. The Lagrangian of (\ref{eq:NZ_problem}) is 
	\begin{equation}\label{eq:apd5}
		\begin{aligned}
	&	\mathcal L^0 = U_t(\bm d)+\nu (g_t - v - \mathbf 1 ^T \bm d) + \bar V_{t+1}(y_t - v;g_t) \\
		& +\underbar{$\lambda$}_v v+ \bar \lambda _v(\min\{\bar v, y_t\} - v) + \underbar{$\boldsymbol{\lambda}$}_d^{T} \bm{d} + \bar 
		{\boldsymbol{\lambda}}_d^{T}(\bar{ \bm{d}} - \bm{d}),
		\end{aligned}
	\end{equation}
	where $\nu$ is the multiplier for the net-zero constraint.
For the deferrable demand, the first order conditions with respect to $v$ on (\ref{eq:apd5}) take the same form as (\ref{eq:foc1})-(\ref{eq:foc2}) with $\nu$ replacing $\pi_t^+$. For $\theta_t^0(g_t)$ that satisfies
		\begin{equation}\label{eq:thetazero}
		\partial_y^+ \bar V_{t+1}(\theta_t^0 (g_t); g_t) \le -\nu \le 	\partial_y^- \bar V_{t+1}(\theta_t^0(g_t) ; g_t),
		\end{equation}
		by the monotonicity of $\partial_y^{\pm} \bar V_{t+1}$ and the same edge-case arguments, 
		\[
		v_t^* = \min\{\bar v, [y_t - \theta_t^0(g_t)]^+\}
		\]
		is optimal.

	For the nondeferrable demand, by the first order condition with respect to $\bm d$ applied to (\ref{eq:apd5})
	\[
	d_{it}^* = \min\{\bar d_i, \partial U_{it}^{-1} (\nu)\},
	\]
    which is the optimal solution of 
    \begin{align*}
        \begin{array}{lrl}
        {\bm d}_t^* &= \underset{\{d_{it} \in [0,\bar{d}_i]\}}{\arg\max}  & \sum_{i=1}^K U_{it}(d_{it})\\[1em]
        &\text{\em subject to} & \sum_{i} d_{it} +v^*_t = g_t.
    \end{array}
    \end{align*}
    
    By the monotonicity of the directional derivative for $\bar V_{t+1}$, $\nu$ that satisfies (\ref{eq:thetazero}) is in between $\pi_t^-$ and $\pi_t^+$: $\nu \in [\pi_t^-, \pi_t^+].$ This also implies that for fixed $g_t$, the procrastination parameters satisfy $\theta_t^-(g_t) \le \theta_t^0(g_t) \le \theta_t^+(g_t)$. 
\subsection{Proof of Theorem 1}

Consider the specialization of Theorem~\ref{thm:opt2} with $\bar{\bm d} = 0$, i.e. no nondeferrable demand. Then $d_{it}^* = 0$ in all three net-consumption regions, and the net-consumption zone thresholds in Theorem 2 reduce to
\begin{align*}
\Delta_t^+(g_t, y_t)&= \min \{ \bar v, [y_t - \theta_t^+(g_t)]^+\}, \\
\Delta_t^-(g_t, y_t)&= \min \{ \bar v, [y_t - \theta_t^-(g_t)]^+\}.
\end{align*}
The optimal deferrable load schedule is then:
\begin{equation}\label{eq:apd8}
	v_t^* = \begin{cases}
		\Delta_t^+(g_t, y_t), & g_t < \Delta_t^+(g_t, y_t)\\
		g_t, & \Delta_t^+(g_t, y_t ) \le g_t \le \Delta_t^-(g_t, y_t) \\
		\Delta_t^-(g_t, y_t), & \Delta_t^-(g_t, y_t) < g_t
	\end{cases}
\end{equation}
We verify that (\ref{eq:apd8}) matches equation (\ref{eq:vt}) across the four segments of $y_t$ defined by $\theta_t^-(g_t)$ and $\theta_t^+(g_t)$. Throughout, we use the monotonicity $\theta_t^-(g_t) \le \theta_t^+(g_t)$.

\begin{enumerate}
	\item \underline{Segment $\textcircled{{\scriptsize 0}}$}: $y_t \le \theta_t^-(g_t)$. Both brackets $[y_t - \theta_t^{\pm}(g_t)]^+$ vanish, so $\Delta_t^+ = \Delta_t^- = 0 \le g_t$. By (\ref{eq:apd8}), $v_t^* = \Delta_t^- = 0$, matching (\ref{eq:vt}).
	\item \underline{Segment $\textcircled{{\scriptsize 1}}$}: $\theta_t^-(g_t) \le y_t \le \theta_t^-(g_t) + \min\{\bar v, g_t\}$. Here, $\Delta_t^-(g_t, y_t) = y_t - \theta_t^-(g_t) \le \min\{\bar v, g_t\} \le g_t$, placing $g_t$ in the net-producing region of (\ref{eq:apd8}). Hence, $v_t^* = \Delta_t^-(g_t,y_t) = y_t - \theta_t^-(g_t)$ matching (\ref{eq:vt}). 
	\item \underline{Segment $\textcircled{{\scriptsize 2}}$}: $\theta_t^-(g_t) + \min\{\bar v, g_t\}< y_t \le \theta_t^+(g_t) + \min\{\bar v, g_t\}$. Two sub-cases arise: 
	\begin{itemize}
		\item If $g_t < \bar v$, then $\Delta_t^+(g_t, y_t) = y_t - \theta_t^+(g_t) \le g_t \le \Delta_t^-(g_t, y_t) = y_t - \theta_t^-(g_t)$, so $g_t$ lies in the net-zero region of (\ref{eq:apd8}) and $v_t^* = g_t$. 
		\item If $g_t \ge \bar v$, then, $\Delta_t^-(g_t, y_t) = \bar v \le g_t$, placing $g_t$ in the net-producing region with $v_t^* = \bar v$. 
	\end{itemize}
	Both sub-cases agree with (\ref{eq:vt}), $v_t^* = \min\{ \bar v, g_t\}$.
	\item  \underline{Segment $\textcircled{{\scriptsize 3}}$}: $\theta_t^+(g_t) + \min\{ \bar v, g_t\} \le y_t$. 
	\begin{itemize}
	\item If $g_t < \bar v$, then $\Delta_t^+(g_t, y_t) > g_t$, placing $g_t$ in the net-consuming region with $v_t^* = \min\{\bar v, y_t - \theta_t^+(g_t)\}$. 
	\item If $g_t \ge \bar v$, then, $\Delta_t^- = \Delta_t^+ = \bar v \le g_t$, placing $g_t$ in the net-producing region with $v_t^* = \bar v $. 
\end{itemize}
 \end{enumerate}

\section{Proof of Propositions}\label{sec:appendixpropproof}
\subsection{Proof of Proposition~\ref{prop:1}}
By Lemma~\ref{lemma:optimal_penaltyregime}, it suffices to consider $y_t \le (T-t+1)\bar v$. Let $\bm \pi = (\pi^+, \pi^-)$ denote the time-invariant NEM price vector.  
	
For $t=T$, the Bellman equation (\ref{eq:bellmanequation}) for $\bm x_T = (y_T, g_T)$ is
    \begin{equation}\label{eq:apd7}
        \max_{v \in [0, \min\{\bar v, y_T\}]} \quad 
        -P_{\bm \pi}(v - g_{T}) - q(y_{T} - v).
    \end{equation}
The derivative of the objective with respect to $v$ is $-\pi^- + q'(y_T - v)$ when $v \le g_T$ and $-\pi^+ + q'(y_T - v)$ when $v > g_T$. By Assumption A3, both expressions are positive throughout the feasible set of $v$. Hence, the objective is strictly increasing in $v$, giving $v_T^* = \min\{\bar v, y_T\} = y_T$, which matches (\ref{eq:procras_chargingrate}) for $\theta_T = (T-T)\bar v = 0$.  

   For $t < T$, since the distribution of $g_t$ is independent of actions, it suffices to compare costs on a realization basis. We partition the feasible range of $y_t$ into three segments and show that the policy in (\ref{eq:procras_chargingrate}) is optimal in each.
    
    \begin{enumerate}
    \item \underline{Case 1} $y_t \le \min\{ \bar v, g_t\}$: The entire remaining demand can be served by DG at the effective marginal cost $\pi^-$ (the foregone export revenue). Any choice $v_t < y_t$ defers demand to a later interval, which it must still be served at a marginal cost of at least $\pi^-$ per unit, regardless of future DG realizations. Deferral can therefore only increase the total cost, so $v_t^* = y_t$. 
    \item \underline{Case 2} $\min\{\bar v, g_t\} <y_t \le \theta_t + \min\{\bar v, g_t\}$: Any $v_t > \min\{\bar v, g_t\}$ incurs the purchase cost $\pi^+$ per unit since $P_\pi(v_t - g_t) = \pi^+(v_t - g_t)$. While this reduces future demand, the savings are at most $\pi^+$ per unit, so such purchase cannot strictly improve the total cost. \\
    Conversely, any $v_t < \min\{\bar v, g_t\}$ defers demand, which by the Case 1 argument only increases the total cost. Hence, $v_t^* = \min\{\bar v, g_t\}$.
    \item \underline{Case 3} $\theta_t +\min\{\bar v, g_t\}<y_t \le\theta_t + \bar v$: Here $y_t > \theta_t = (T-t)\bar v$, so serving $v_t < y_t - \theta_t$ fails to complete remaining deferrable demand by its deadline. On the other hand, any $v_t > y_t - \theta_t$ requires purchased power at the cost $\pi^+$ per unit, which by the argument in Case 2, it only increases the total cost. Hence, $v_t^*  = y_t - \theta_t$. 
    \end{enumerate}
	In all three cases, the policy in (\ref{eq:procras_chargingrate}) achieves the minimum realization-wise cost. Since this holds for every sample path and every $t$, Proposition~\ref{prop:1} holds for all $t$.
\subsection{Proof of Proposition 2}
	We prove $\theta_t^+ = \theta_{t+1}^+ + \bar v$; the argument for $\theta_t^- = \theta_{t+1}^-$ is parallel and sketched at the end. 
	
	We verify the first order condition (\ref{eq:thetaplus}) at $y = \theta_{t+1}^+ + \bar v$ for time $t$. By (\ref{eq:thetaplus}), $\theta_{t+1}^+$ satisfies,
    \begin{equation*}
    	\partial _y ^+  \bar V_{t+2}(\theta_{t+1}^+) \le -\pi_{t+1}^+ \le	 \partial _y ^-  \bar V_{t+2}(\theta_{t+1}^+).
    \end{equation*}
By Lemma~\ref{lemma:concavity}, $V_t$ is concave in $y$, so the directional derivatives $\partial_y^{\pm}\bar V_t$ are non-increasing. It suffices to show
	\begin{equation}
        	\partial _y ^+  \bar V_{t+1}(\theta_{t+1}^+ + \bar v) \le -\pi_{t}^+ \le	 \partial _y ^-  \bar V_{t+1}(\theta_{t+1}^+ + \bar v) .
	\end{equation}
	For $y_{t+1} \ge \theta_{t+1}^+ + \bar v$, the monotonicity $\theta_{t+1}^- \le \theta_{t+1}^0 \le \theta_{t+1}^+$ implies, via Theorem~\ref{thm:opt2}, that $v_{t+1}^* =\bar v$ for all three net-consumption regions. Hence, for all $g_{t+1}$ and $y_{t+1} \ge \theta_{t+1}^+ + \bar v$,
	 \begin{equation}\label{eq:34}
	 	\begin{aligned}
		V_{t+1}&(y_{t+1} , g_{t+1}) = U(\bm d_{t+1}^*) \\
		&- P_{\pibf_{t+1}}(\mathbf 1^T \bm d_{t+1}^* + \bar v - g_{t+1})  
		+ \bar V_{t+2}(y_{t+1} - \bar v).        
		\end{aligned}
	\end{equation}
	Then, the right derivative of $\bar V_{t+1}$ at $y_{t+1} = \theta_{t+1}^+ +  \bar v$ satisfies 
	\[
	    \partial_y^+ \bar V_{t+1}(\theta_{t+1}^+ + \bar v) = \partial_y ^+ \bar V_{t+2}(\theta_{t+1}^+) \le -\pi_{t+1}^+ = -\pi_{t}^+. 
	\]
 For $\theta_t^+ \le y_t \le \theta_t^+ + \bar v$, by Theorem~\ref{thm:opt2}, the optimal charging rate in the net-consuming region is $v_{t+1}^* = y_{t+1} - \theta_{t+1}^+$, yielding the remaining demand $y_{t+2} = \theta_{t+1}^+$. In the net-producing and net zero regions, the monotonicity of the procrastination parameters implies $v_{t+1}^* \ge y_{t+1} - \theta_{t+1}^+$, so the remaining demand satisfies $y_{t+2} \le \theta_{t+1}^+$, for all three regions' optimal decisions.

    The left derivative of $\bar V_{t+1}(y_{t+1})$ at $y_{t+1} = \theta_{t+1}^+ + \bar v$ is:
    \begin{align*}
    &	\partial_y^- \bar V_{t+1}(y_{t+1}) = \Pr(\text{Net-producing at } t+1) \partial _y ^-  \bar V_{t+2}(y_{t+2}^-)  \\ 
    	&+\Pr(\text{Net-consuming at } t+1) \partial _y ^- \bar V_{t+2}(\theta_{t+1}^+) \\
    	&+ \Pr(\text{Net-zero at }t+1) \mathbb E[\partial _y^- \bar V_{t+2}(y_{t+2}^0) | \text{Net-zero at }t+1],
    \end{align*} 
    where $y_{t+2}^-$ and $y_{t+2}^0$ are resulting remaining demand conditioning on $t+1$ is net-producing and net-zero regions, respectively. Since $y_{t+2}^-, y_{t+2}^0 \le \theta_{t+1}^+$ and $\partial_y^-\bar V_{t+2}$ is non-increasing, each conditional derivative satisfies $\partial_y^-\bar V_{t+2}(\cdot)\ge \partial_y^-\bar V_{t+2}(\theta_{t+1}^+) \ge -\pi_{t+1}^+$. Therefore, 
    \[
    \partial_y^- \bar V_{t+1}(\theta_{t+1}^+ + \bar v) \ge -\pi_{t+1}^+ = -\pi_t^+.
    \]
    The proof of $\theta_t^- = \theta_{t+1}^-$ follows by an analogous argument: replace $\pi_t^+$ by $\pi_t^-$, and evaluate the directional derivative of $\bar V_{t+1}$ at $y_{t+1} = \theta_{t+1}^-$. The monotonicity of directional derivatives and three net-consumption regions decomposition yields the required bounds (\ref{eq:thetaminus}).

\section{Sensitivity Analysis}\label{sec:appendixC}
We present sensitivity analysis of PTRL for learning rate and hidden layer size. W

\subsection{Learning rate sensitivity analysis}

\begin{figure}[t]
	\begin{subfigure}[b]{0.49\columnwidth}
		\includegraphics[width=\linewidth]{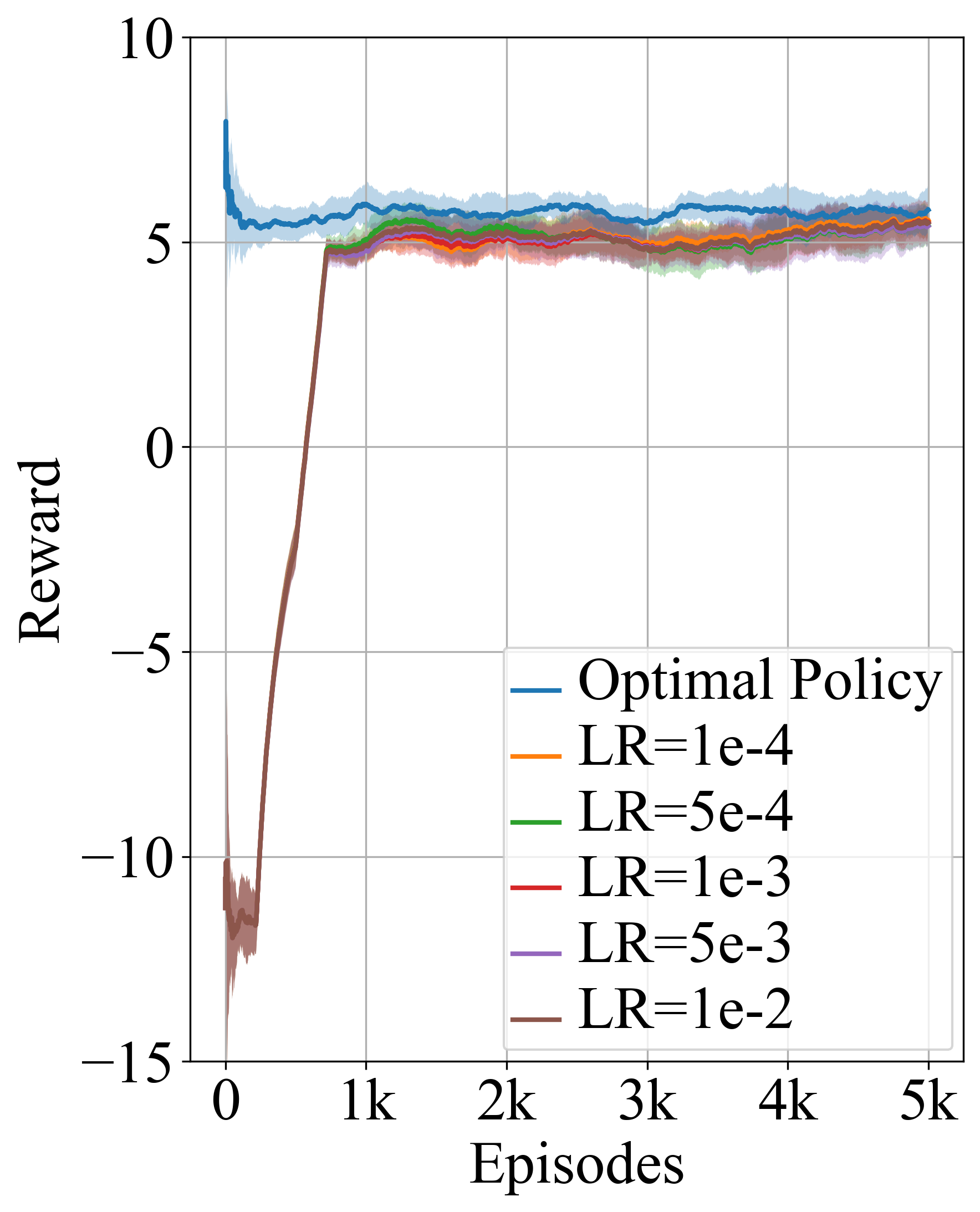}
		\caption{}
		\label{fig:sensitivity_LR}
	\end{subfigure}
	\hfill 
	\begin{subfigure}[b]{0.49\columnwidth}
		\includegraphics[width=\linewidth]{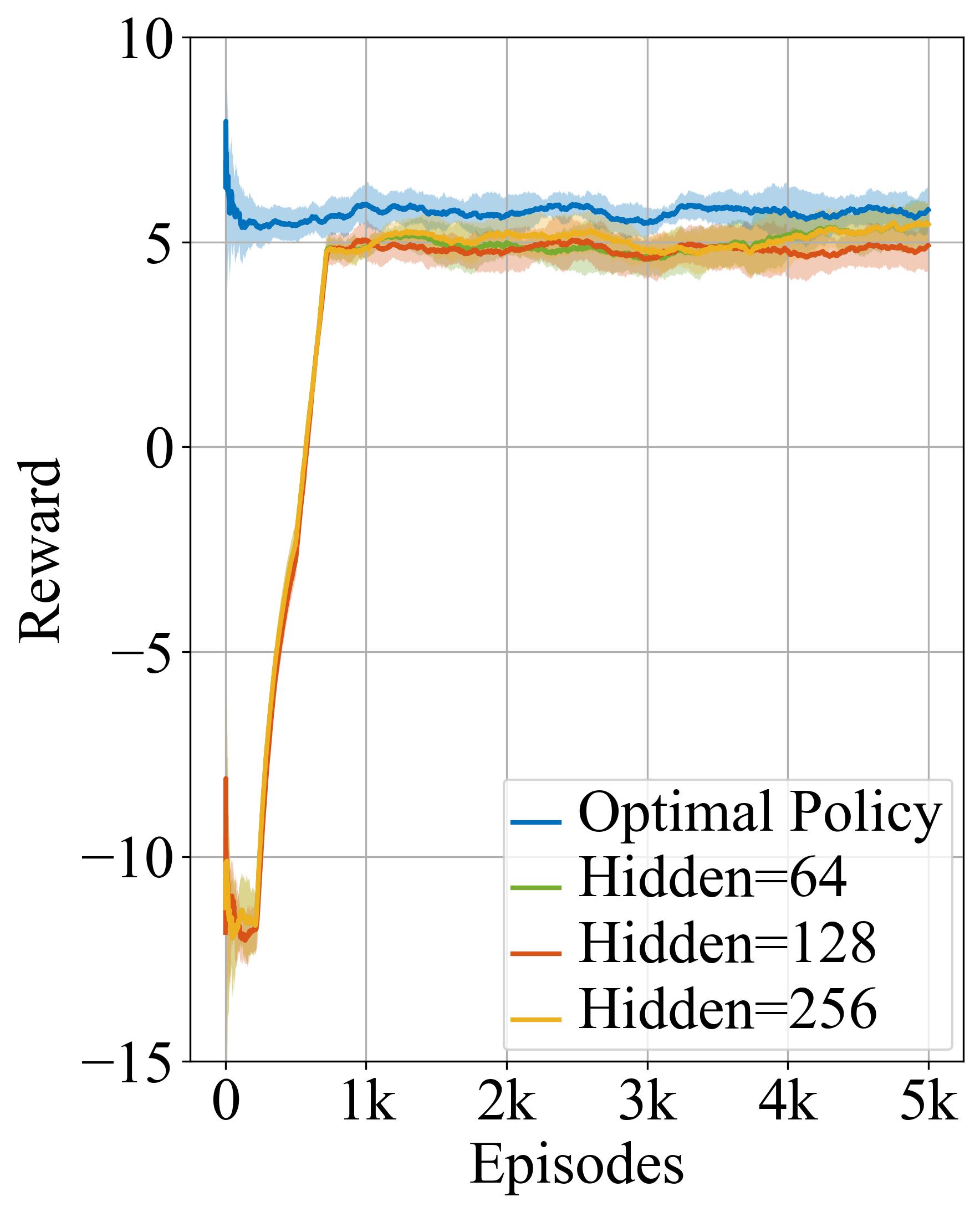}
		\caption{}
		\label{fig:sensitivity_hidden}
	\end{subfigure}
	\caption{Sensitivity analysis of PTRL for (a) learning rate and (b) hidden layer size}
	\label{fig:sensitivityanalysis}
\end{figure}

Fig~\ref{fig:sensitivity_LR} shows the learning curves of PTRL under five learning rates ranging over two orders of magnitude, $\lambda \in \{10^{-4}, 5 \times 10^{-4}, 10^{-3}, 5 \times 10^{-3}, 10^-2\}$, with the actor and crittic sharing the same rate. The setup is otherwise identical to Sec.~\ref{sec:VB} under the time-independent DG model, with each curve averaged over the same 10 random seeds used in Fig.~\ref{fig:LearningCurve}. All five settings exhibit nearly identical learning behavior: an initial transient in the first few hundred episodes, followed by convergence at approximately 700–800 episodes, with no monotonic trend in convergence speed across $\lambda$.  The final cumulative rewards at the end of training were 5.53, 5.40, 5.49, 5.39, and 5.47 in increasing order of $\lambda$, corresponding to gap-to-optimality values of 4.41\%, 6.62\%, 5.15\%, 6.81\%, and 5.45\% relative to the optimal cumulative reward of 5.79. The normalized standard deviations across seeds were 0.066, 0.088, 0.089, 0.091, and 0.064, all comparable in magnitude and well within the across-seed variation observed for PTRL in Fig.~\ref{fig:LearningCurveIndependent}. The differences across learning rates are within seed noise, indicating that PTRL is robust to the choice of learning rate over two orders of magnitude. We attribute this robustness to the dimension reduction afforded by the procrastination structure: optimization over the three-dimensional threshold space is sufficiently well-conditioned that the algorithm tolerates a wide range of step sizes without destabilization.

\subsection{Hidden layer size sensitivity analysis}
Fig. 13(b) shows the learning curves of PTRL under three hidden layer sizes, with both layers of the actor and critic MLPs set to 64, 128, and 256 neurons. All other settings follow Sec.~\ref{sec:VB} under the time-independent DG model, with results averaged over the same 10 random seeds. The convergence behavior is again nearly indistinguishable across configurations, with all three settings stabilizing at approximately 700–800 episodes. The final cumulative rewards were 5.44, 4.92, and 5.44 in increasing order of hidden layer size, corresponding to gap-to-optimality values of 5.93\%, 14.99\%, and 5.93\% relative to the optimal cumulative reward of 5.79. The normalized standard deviations across seeds were 0.098, 0.126, and 0.093, all comparable in magnitude. These results indicate that PTRL is robust to the choice of hidden layer size, again reflecting the benefit of optimizing over the low-dimensional threshold space rather than the full $(K+1)$-dimensional action space. Among the three settings, the smallest network (64 neurons per layer) achieves performance on par with the largest while requiring substantially fewer parameters, suggesting that a smaller hidden layer size may be preferred in practice.

\bibliography{ref.bib}
\bibliographystyle{IEEEtran}

\end{document}